\documentclass[fleqn,usenatbib]{aa}

\DeclareRobustCommand{\VAN}[3]{#2}
\let\VANthebibliography\thebibliography
\def\thebibliography{\DeclareRobustCommand{\VAN}[3]{##3}\VANthebibliography}

\usepackage{amsmath}	
\usepackage{amssymb}	
\usepackage{epstopdf}
\usepackage{threeparttable}
\usepackage{bm}
\usepackage{soul}
\usepackage{blindtext}
\usepackage{times}
\usepackage{natbib}

\usepackage{graphicx}
\usepackage{txfonts}
\usepackage{natbib}
\usepackage{hyperref} 

\newcommand{\eqb}{\begin{eqnarray}}
\newcommand{\eqe}{\end{eqnarray}}

\newcommand{\oiiil}{[O\,{\footnotesize III}]$\lambda$5007}
\newcommand{\oiii}{[O\,{\footnotesize III]}}
\newcommand{\oii}{[O\,{\footnotesize II]}$\lambda$3727}

\newcommand{\oiil}{[O\,{\footnotesize II]}}

\newcommand{\hb}{\rm H\ensuremath{\beta}}

\newcommand{\niir}{[N\,{\footnotesize II}] $\lambda$6584}

\newcommand{\sar}{\lambda_\mathrm{sBHAR}}
\newcommand{\Sar}{$\sar$}

\def\lephare{\texttt{LePhare}}
\def\classic{\textsc{Classic}}
\def\farmer{\textsc{The Farmer}}
\def\tractor{\textsc{The Tractor}}

\begin{document}

\title{Obscured and unobscured X-ray AGNs I: Host galaxy properties}

\author{
C. G. Bornancini\inst{1,2},
G. A. Oio\inst{2,3}, G. Coldwell\inst{4}} 
\institute{Instituto de Astronom\'{\i}a Te\'orica y Experimental, (IATE, CONICET-UNC), C\'ordoba, Argentina \\
\email{cbornancini@unc.edu.ar}
\and
Universidad Nacional de C\'ordoba, Observatorio Astron\'omico de C\'ordoba, Laprida 854, X5000BGR, C\'ordoba, Argentina\\
\and
Consejo de Investigaciones Cient\'ificas y T\'ecnicas de la Rep\'ublica Argentina – CONICET, Godoy Cruz 2290, C1425FQB, CABA, Argentina
\and 
Facultad de Ciencias Exactas, F\'isicas y Naturales, Departamento de Geof\'isica y Astronom\'ia, CONICET Universidad Nacional de San Juan, Av. Ignacio de la
Roza 590 (O), J5402DCS Rivadavia, San Juan, Argentina
}

\date{Received XXX; accepted YYY}

\abstract
{Active galactic nuclei (AGN) play a crucial role in galaxy evolution by influencing the observational properties of their host galaxies.} 
{We investigate the host galaxy properties of X-ray selected AGNs, focusing on differences between obscured and unobscured AGNs, and between high-(log(\oiiil/\hb$\ge$0.5) and low-excitation sources (log(\oiiil/\hb$<$0.5).}
{We selected a sample of AGNs from the spectroscopic zCOSMOS survey with $0.5 \leq z_{\rm sp} \leq 0.9$ based on the Mass-Excitation (MEx) diagram and X-ray emission. AGNs were classified as obscured or unobscured using hydrogen column density, and as high- or low-excitation based on the \oiiil/\hb~ratio.
We analysed various AGN properties, including the hardness ratio, X-ray luminosity, emission line ratios such as the ionisation-level sensitive parameter O32=log(\oiiil/\oii), and the metallicity sensitive parameter R23=log((\oiiil+\oii)/\hb), and the specific black hole accretion rate (\Sar). 
}
{Unobscured AGNs exhibit a more evident correlation between the \oiiil/\oii~ionisation ratio and X-ray luminosity than obscured AGNs, while high-excitation obscured AGNs reach, on average, higher X-ray luminosities. Furthermore, high-excitation AGNs typically show high values of R23, suggesting low metallicities, similar to that observed in high-redshift galaxies ($4< z <6$). 
We find a positive correlation between the parameters $\lambda_\mathrm{sBHAR}$ and $N_{\rm H}$, R23 and O32 parameters. The correlation suggests that AGNs with a high specific accretion rate have not only a higher production of high-energy photons, which ionise the surrounding medium more intensely, but are also usually associated with environments less enriched in heavy elements. 
These results provide insights into the complex interplay between AGN activity, host galaxy properties, and the role of obscuration in shaping galaxy evolution.
}
{}

\titlerunning{Properties of obscured and unobscured AGNs}
\maketitle
\section{Introduction}

Active galactic nuclei (AGNs) are one of the most interesting objects to study the formation and evolution of galaxies \citep{miley08, netzer2015, hickox2018}. These consist of small regions located at the centre of galaxies that emit an extraordinary amount of energy by non-thermal processes, i.e. that are not related to the emission of energy from stars \citep{ambart58}.
After the discovery of quasars (QSOs) in the 1960s \citep{sch}, the most luminous counterpart of AGNs, various studies have been carried out on the nature of these objects.
Nowadays it is well accepted that the emission of energy by AGNs is due to the accretion of matter (gas) by a supermassive black hole at the centre of its host galaxy \citep{Lynden1969, lynden1971, rees1974, rees1984}. 

AGN classification has historically been based on spectral features, distinguishing between Type I and Type II AGNs. According to the Unified Model \citep[UM,][]{anto93, urry95,netzer2015}, this distinction arises primarily from orientation effects, where a dusty torus surrounding the accretion disk obscures the central engine along certain lines of sight. 

In light of the UM's predictions, it is expected that the properties of the host galaxies will be comparable, given that the observed discrepancies would be attributable to the orientation of the line of sight in the innermost parts of the galaxy nucleus \citep{Zou19}. However, some studies show discrepancies in the sizes of these central structures \citep{alonso-herrero11,ramos2011, martinez2017,audibert17}, suggesting that the shape of the torus may not be as regular as previously thought. This irregularity could be caused by larger-scale material, rather than parsec-scale tori \citep{maiolino95, gould12, donley18,silverman23}. Alternatively, even the torus or the BLR might not be stable and could disappear within a relatively short time frame \citep{Elitzur2006, Elitzur2009}.
 
In recent years, a model has been developed which considers the various stages of galaxy formation, based mainly on results obtained in cosmological simulations. In this evolutionary scenario, the earliest phases of galaxy formation would be dominated by violent interactions, mergers, or strong instabilities between small galaxy systems. This would result in strong star formation episodes, where gas loses angular momentum and moves towards the central region, feeding the central SMBH and forming an obscured nucleus \citep{dimatteo2005,sanders07, hopkins08a, hopkins08b, springel2005, springel2005Natur,alexander2012}. During these stages, the growth of SMBHs would result in the release of significant energy in the form of radiation, outflows, and relativistic jets \citep{hickox2018}. The bright galactic nucleus and supernova-driven winds would produce strong feedback that would sweep the gas, resulting in a very bright optical QSO \citep{springel2005}. As evolution progresses, the star formation and accretion rates will decrease, resulting in a spheroid with a passive evolution. 

It is widely accepted that there are correlations between different macro-parameters of the host galaxies and those belonging to the central microstructure of the AGNs. However, it should be noted that the relative dimensions of these parameters are completely different and of several orders of magnitude. For example, there is a correlation between the mass of the bulge and the mass of the central hole \citep{magorrian98, wandel, mclure,haring04, graham15,ding}, between black hole mass and the velocity dispersion \citep{ferrarese00,Gebhardt00, merrit01,beifiori, mcconnel13}. The identified correlations suggest that the evolution of the galaxy and its central SMBH are linked, supporting the theoretical premise that the formation of spheroids and the growth of supermassive black holes are driven by a common process \citep{granato2004,Ho2008,kormendy_Ho2013,heckman2014,
brandt2015,padovani2017,ramos2017,lopez23}.

Recently, several techniques have emerged for the identification of galaxies with AGNs. These include mid-IR selection techniques \citep{L04, S05, L07, donley07, donley08, assef2011, M12, Ch17}, diagnostic diagrams based on optical emission line ratios \citep{J11, J14, Zh18} and according to X-ray emission \citep{Mushotzky2004, hasinger2005, Dewangan2008}. 
The selection of AGNs with X-ray emission presents great advantages over those selected in optical or in the IR wavelengths \citep{bornan2020, bornan2022}. X-ray emission from AGNs is thought to originate on small scales close to the SMBH \citep{Buchner2014} and can be used to identify AGNs directly. High-energy X-rays selection is less biased by obscuration effects by stellar light from the host galaxy \citep{comastri2002, Georgantopoulos2005} and can penetrate through large gas column densities, demonstrating that this method is the most effective, reliable and complete for selecting AGN samples \citep{Mushotzky2004,brandt2015}. 

A key aspect in the study of AGNs is the analysis of emission lines, which provide direct information on the physical and chemical conditions of the ionised gas. One of the most relevant spectral diagnostics is the ratio of oxygen lines, such as the ratio \oiiil/\oii (hereafter O32). This ratio is a critical indicator of gas ionisation, being sensitive to both the hardness of the AGN radiation field and the metallicity of the surrounding medium. Highly excited AGN show high values of \oiiil/\oii, reflecting an environment ionised by extremely energetic radiation \citep{bassett, paalvast18}.
Furthermore, the parameter R23, defined as R23= \oiiil+\oii)/\hb~\citep{pagel79, kewley02, tremonti04, Nakajima}, is a valuable tracer of the abundance of oxygen in the galaxy. R23 is commonly used as an indicator of the metallicity of the ionised gas, in conjunction with other line ratios, to map the metal distribution in AGN and active galaxies. Measuring R23 provides an indirect way to assess the degree of chemical enrichment of the ionised gas by the AGN, which in turn is related to stellar evolution and the feedback process caused by the active nucleus.
By combining parameters such as O32 and R23, we can investigate the physical conditions of the ionised gas in AGN gain insights into the evolutionary history of the host galaxies. This enables us to comprehend the impact of these active nuclei on the surrounding interstellar medium and star formation \citep{nakajima13, chemo}. The study of these spectral ratios remains a fundamental tool for understanding the nature and evolution of AGN over cosmic time \citep{remetallica}.

The study of the specific black hole accretion rate $\lambda_\mathrm{sBHAR}$ is crucial to understanding the impact of AGNs on their host galaxies. In particular, it can elucidate the role of energetic feedback in regulating star formation \citep{Aird18}. A high $\lambda_\mathrm{sBHAR}$ implies an active and efficient accretion, which can be associated with powerful winds, X-ray emission and other forms of feedback that impact the galactic environment \citep{mountricas_geor}. However, low $\lambda_\mathrm{sBHAR}$ values indicate a phase of low activity or "quiescent AGN".
The relationship between $\lambda_\mathrm{sBHAR}$ and other galactic properties, such as the hydrogen column density, has also been the subject of recent studies. These studies have explored how galaxies, with obscured and unobscured cores, differ in their specific accretion rates. The results of research conducted by \citet{aird12} and \citet{kauff09} suggest that AGNs with high $\lambda_\mathrm{sBHAR}$ are typically with galaxies experiencing active star formation and low metallicity. Conversely, those with low $\lambda_\mathrm{sBHAR}$ may be in more evolved stages, with a limited supply of gas for accretion \citep{aird12, mountrichas24}.

In this paper we analyse the properties of galaxies hosting AGNs. The data was primarily obtained through the ratio of specific spectral lines, identified through optical spectroscopy and X-ray photometry. To achieve this, we will use line diagnostic diagrams utilising the ratio of \oiii/\hb\ and the host galaxy stellar mass \citep{J14}.

This paper is organised as follows: 
 In Section~\ref{data}, we present all datasets used in this study and in Section~\ref{selection} we detail the AGN selection methods. In Section~\ref{properties}, we analyse the main properties of AGNs such as X-ray luminosity, star formation rates, and specific black hole accretion rates. Also parameters that allow us to investigate the obscuration such as the hardness ratio and the hydrogen column density as well as line quotient of \oiiil, \oii and \hb~ lines. And finally, the summary and discussion of our study are presented in Section~\ref{sum}.

Throughout this work we will use the AB magnitude system \citep{oke} and we will assume a $\Lambda$CDM cosmology with H$_{0} = 70$ km s$^{-1}$ Mpc$^{-1}$, $\Omega_{\rm M} = 0.3$, $\Omega_{\rm \Lambda} = 0.7$.

\section{Datasets}
\label{data}
To investigate the properties of X-ray-selected AGNs, we use data obtained from the Cosmic Evolution Survey (COSMOS) \citep{scoville}. COSMOS survey is a deep, wide area and multi-wavelength observational project of about two square degrees and nearly 2 million galaxies that compresses images obtained in optical, UV, IR, by terrestrial (Very Large Telescope, Subaru Telescope, Canada France Hawaii Telescope) and space telescopes (Spitzer, Herschel, GALEX, HST, JWST).
Particularly we utilise the galaxy catalogues known as COSMOS2015 \citep{laigle} and its latest version/update COSMOS2020 \citep{weaver} as well as spectroscopic data obtained in the zCOSMOS catalogue \citep{lilly07,lilly09} and X-ray data \citep{civano, Laloux}.

The COSMOS2015 catalogue\footnote{The catalogue can be downloaded from \url{ftp://ftp.iap.fr/pub/from\_users/hjmcc/COSMOS2015/}}  \citep{laigle} is a multi-band catalogue with photometry covering a broad range of wavelengths, from the X-ray range through to the radio of half a million galaxies. Furthermore, the catalogue contains measurements of photometric redshifts calculated using \lephare{} \citep{arnouts11} as well as the galaxy stellar mass, the star formation rate (SFR) and the specific star formation rate (sSFR=SF/M$_*$).
 
The COSMOS2020 catalogue\footnote{\url{https://cosmos2020.calet.org/}} is an update of the catalogue published in 2015. It includes new IR photometry data and new determinations of photometric redshifts and other galaxy parameters \citep{weaver}. These authors present two catalogues created using two different methods: \farmer{} and \classic{} catalogues. The \classic{} was developed using standard techniques, as outlined by \citet{laigle}. These techniques include aperture photometry performed on PSF-homogenised images and the use of IRACLEAN software \citep{Hsieh2012} for photometry in IRAC images. The \farmer{} catalogue was created to perform profile-fitting photometry using the \tractor{} code. This derives entirely parametric models from one or more images and provides also morphological information \citep{Lang2016}.
 
zCOSMOS is a Large Program on the European Southern Observatory (ESO) Very Large Telescope (VLT) comprising 600 hours of observation used to carry out a major redshift survey with the Visible Multi Object Spectrograph (VIMOS). 
This redshift survey is divided into two parts. The first part comprises spectra for approximately 28.000 galaxies at 0.2 $< z <$ 1.2 selected to have I$_{AB}$ $<$ 22.5. This is known as zCOSMOS-bright, which covers an approximate area 
of 1.7 deg$^2$ of the COSMOS field  with a high success rate ($\sim$ 70\%) in measuring redshifts. It achieves a success rate of 100\% at 0.5$ < z < $0.8 \citep{knobel12}. The second part, known as zCOSMOS deep, contains approximately 12.000 galaxies at 1.2$ < z <$ 3 with I$_{AB}$ $<$ 22.5 chosen by two colour-selection criteria ($B-Z$) vs. ($Z-K$) and ($U-B$) vs. ($V-R$) at a sampling rate of 70\%.

Despite the existence of several spectroscopic surveys in this area such as DEIMOS 10k \citep{hasinger18}, MUSE Wide survey \citep{urrutia19}, FMOS-COSMOS \citep{silverman15,kashino19} and the MOSFIRE Deep Evolution Field Survey  \citep{kriek15}, we have only used the zCOSMOS-bright catalogue\footnote{https://cdsarc.cds.unistra.fr/ftp/J/ApJS/172/70/zcosmos3.dat.gz} due to two particular reasons: because it represents a homogeneous catalogue using the same selection criteria and with a very good completeness, secondly, we have used spectral line measurements obtained from the ASPIC 
(Archive of Spectrophotometry Publicly available In Cesam) database.
The ASPIC database is dedicated to providing access to added-value data produced from a number of public spectroscopic or spectrophotometric surveys. 
This catalogue\footnote{The catalogue can be downloaded from \url{https://cesam.lam.fr/zCosmos/search/download}} 
 collates line flux and equivalent width measurements for all the zCOSMOS sources using two independent software packages: \textsc{platefit vimos} \citep{lama2006} and slinefit \citep{Schreiber2018}. 
Spectroscopic measurements (emission and absorption lines
fluxes, and equivalent widths) in zCOSMOS were performed
with the automated pipeline \textsc{platefit vimos} \citep{lama2006, lama2009}.
 This routine removes the stellar continuum and absorption lines and then fits automatically all emission lines as a combination of 30 single stellar population (SSP) templates, with different ages and metallicities from the \citet{Bruzual2003} library. The best-fit synthetic spectrum is used to remove the stellar component. Subsequently, the emission lines are fitted as a single nebular spectrum, comprising a sum of Gaussians at specified wavelengths.

\section{AGN selection}
\label{selection}

\begin{figure*}
	\includegraphics[width=85mm]{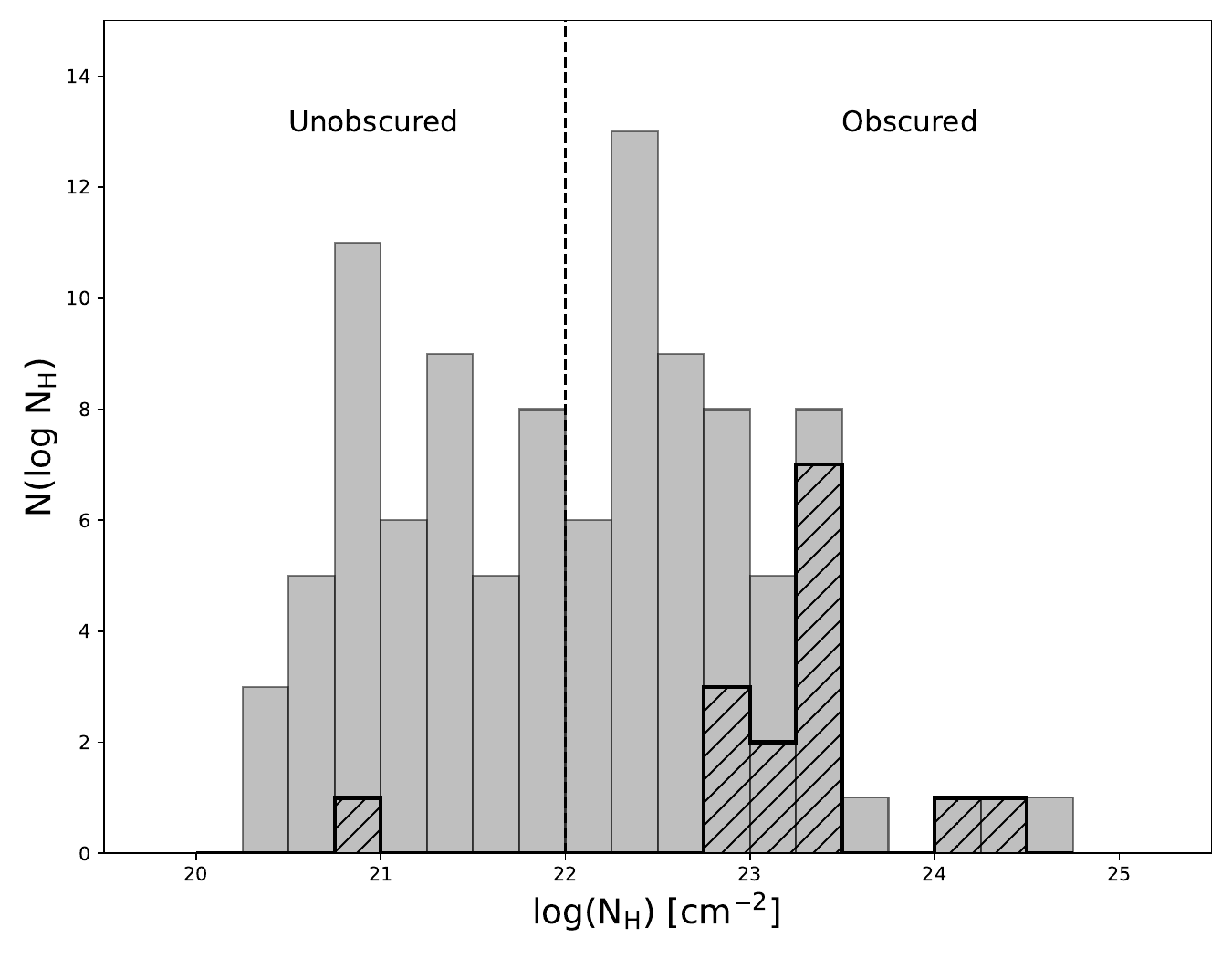}
 	\includegraphics[width=85mm]{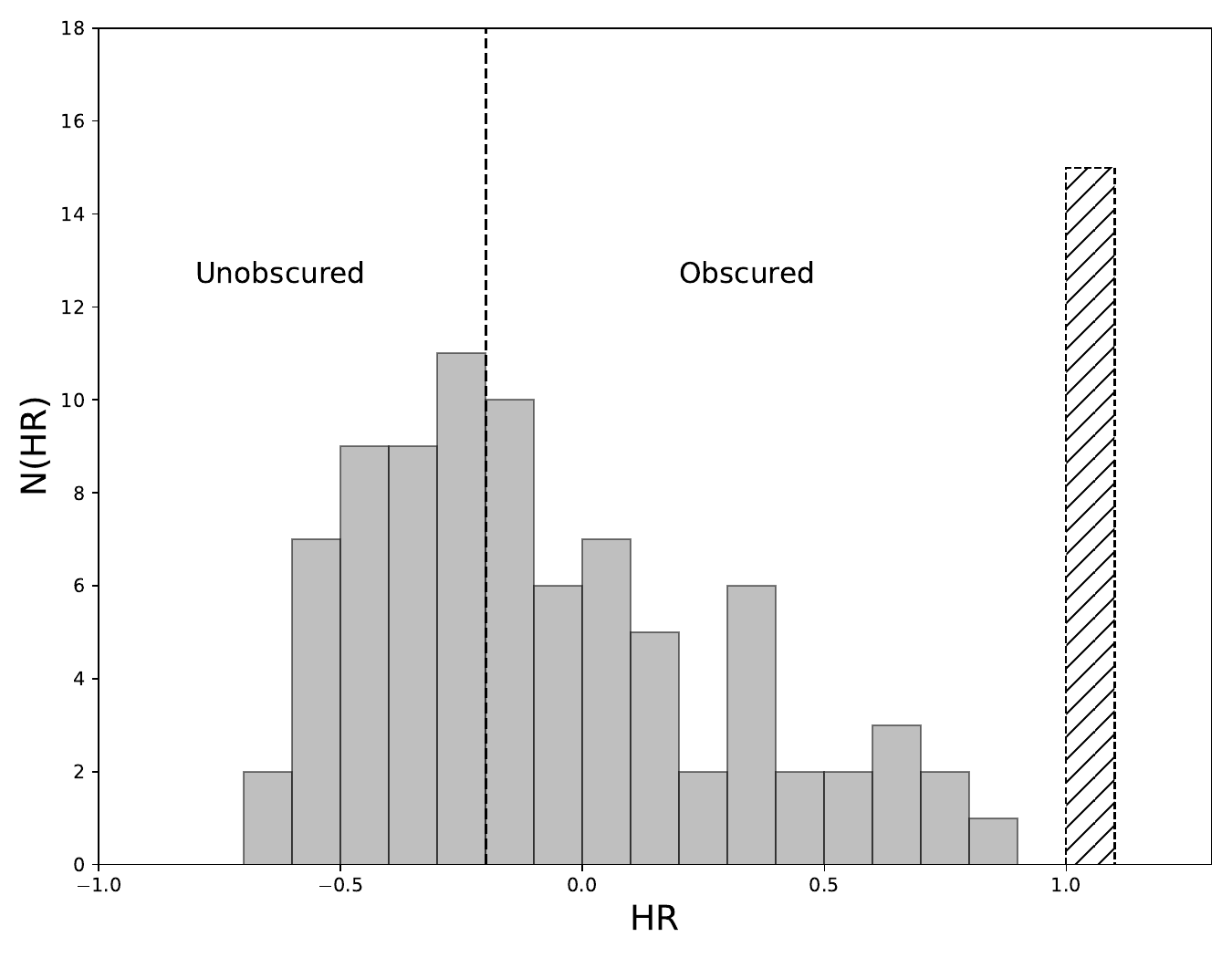}
    \caption{Distribution of neutral hydrogen column density log($N_{\rm H}$) (left panel) and hardness ratios HR (right panel) for AGNs selected using the MEx diagram. Objects that do not show absorption have been plotted at HR=1, and are represented with a solid line histogram in the left panel. The dotted vertical lines at log($N_{\rm H}$)=22 and HR=$-$0.2 indicate the adopted dividing lines between obscured and unobscured AGNs.}
    \label{nh}
\end{figure*}

To select the AGN sample, we have used a line diagnostic method proposed by \citet{J11, J14}. This method, known as the Mass-Excitation diagnostic diagram (MEx), is used to classify and separate AGNs from star-forming and composite galaxies, employing the ratios of the \oiiil~ and \hb~lines and the galaxy stellar mass.
This selection criterion is similar to the traditional Baldwin-Phillips-Terlevich (BPT, \citet{bpt}) method but with the exception of using the stellar mass instead of the quotient \niir/H$\alpha$. This is due to the fact that the quotient moves to the near-IR wavelengths for $z > 0.4$. 
In this work we employ the criterion of \citet{J14} which introduces small modifications to the original method presented in \citep{J11}. 

In order to obtain an AGN sample selected with an optimal signal-to-noise ratio (S/N), we have analysed some results from previous studies. \citet{bongiorno2010} conducted an analysis of the luminosity function of AGN selected from \oiiil~ line emission applied a cut to the signal-to-noise ratio of S/N > 5 calculated using the \oiiil~line, while other lines were selected using S/N > 2.5. \citet{caputi2013} studied the optical spectrum of sources selected at 24 $\mu$m from the zCOMOS-bright catalogue finding, for AGNs, that the most secure measurements were for lines with EW$>$5 \AA. 
\citet{marziani2017} studied a sample of emission lines from galaxy members in the WIde-field Nearby Galaxy cluster Survey (WINGS). They found that the S/N ratio of the \oiiil~line was representative of the overall S/N of each spectrum, finding that a S/N$\approx $6 would correspond to a minimum $EWmin\approx$3 \AA. Similarly, \citet{kong2002} found that AGNs with EW$>$1.5 \AA~have 3$\sigma$ detection levels in a spectroscopic study of blue compact galaxies.
Following these results, we have selected objects with EW$ > $3 \AA~on the \oiiil, \oii~and \hb~ lines which corresponds to S/N $\geq$6. The aim is to utilise these emission line measurements to examine the ionisation-level sensitivity \oiii/\oiil~ratio and a parameter associated with the metallicity of galaxies, the later of which will be developed in greater detail in Section \ref{ioni}. 

As previously stated, X-ray emission is employed to guarantee the reliability of the AGN sample and to investigate the properties of obscured and unobscured X-ray selected AGNs. In order to achieve this, we have correlated our galaxy catalogue with the X-ray sample taken from the COSMOS-Legacy Survey (CLS) catalogue \citep{civano}. This contains measurements of the hardness-ratio, calculated as HR=H-S/H+S, where $H$ and $S$ are the count rates in the hard (2–10 keV) and soft bands (0.5–2 keV), respectively. To obtain the neutral hydrogen column density $N_{\rm H}$ our catalogues have been correlated with an X-ray catalogue taken from \citet{Laloux}. In this catalogue, the $N_{\rm H}$ measurements were calculated using the Bayesian X-ray Analysis (BXA) package presented by \citet{Buchner2014} and the UXCLUMPY torus model presented in \citet{buchner2019}. For further details, the reader can consult the work of \citet{Laloux}.

\begin{figure}
	\includegraphics[width=90mm]{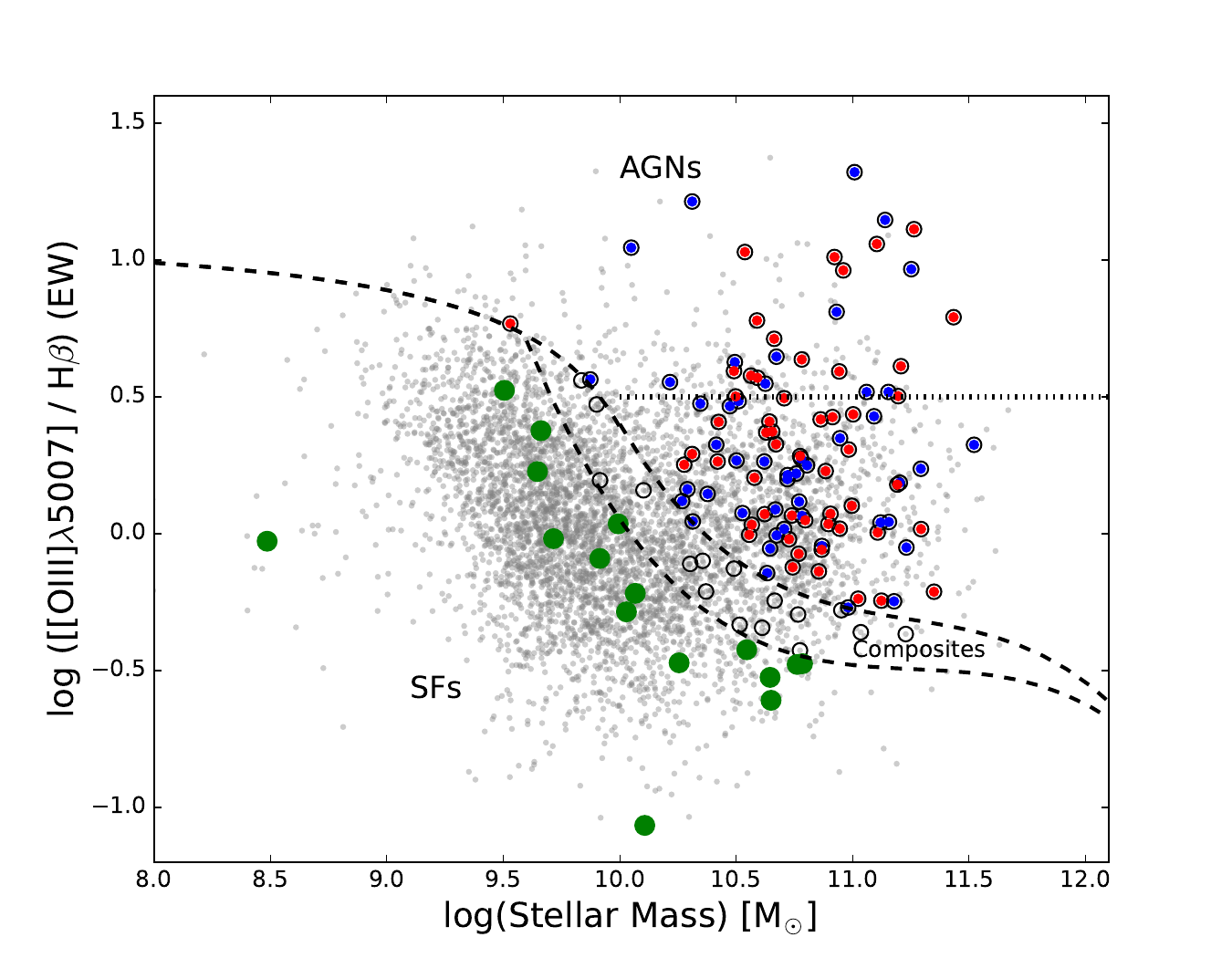}
         \caption{The Mass-Excitation diagram. Grey dots represent galaxies in the zCOSMOS survey with ASPIC spectral line measurements.         
         The regions marked with dashed lines show the location of AGNs, composites and star-forming galaxies. Grey circles represent X-ray sources and red and blue circles represent obscured and unobscured AGNs with X-ray emission. Green circles denote X-ray detected sources identified as star-forming galaxies. Dot horizontal line at $y=0.5$ shows the separation between high and low excitation AGNs. }
    \label{mex}
\end{figure}

In Figure \ref{nh} we show the log($N_{\rm H}$) (left panel) and HR (right panel) distributions for AGNs selected using the MEx diagram (see Figure \ref{mex}). 
Vertical dashed lines in both panels show the usual criterion used to separate obscured and unobscured AGNs in the X-rays.
In the left panel of this figure the dashed vertical line shows the  log($N_{\rm H}$) value (log($N_{\rm H}$)$=22$) which is used by several authors \citep{civano2012} to separate obscured and unobscured sources in the X-rays at all redshifts. Right panel shows the typical hardness ratio value used in the bibliography to separate obscured and unobscured sources \citep{sanchez17}. 
The corresponding limits are at log($N_{\rm H}$)=22/cm$^{-2}$ \citep{buchner2015,Koutoulidis2018} and HR=$-$0.2 \citep{gilli2005, treister09, marchesi}.
We have noticed 15 AGNs that lack HR measurements (shown with HR$=$1). These correspond mainly to obscured AGNs according to the distribution of log($N_{\rm H}$) in the left panel of Figure \ref{nh} (solid line histogram).
To assess the consistency between the obscuration indicators, we compared the classifications based on hydrogen column density ($\log N_{\mathrm{H}}$) and hardness ratio (HR). Among obscured sources ($\log N_{\mathrm{H}} > 22$), 65\% have HR~$>0.2$, while 94\% of the unobscured sources (log $N_{\mathrm{H}} \leq 22$) have HR~$\leq 0.2$. This indicates a good overall agreement, with discrepancies likely due to redshift effects, source geometry, or the limitations of HR as a proxy for absorption. For this reason, we have decided to use the log($N_{\rm H}$) values as a separation criterion between obscured and unobscured AGNs. 
The final sample consists of 53 and 47 obscured and unobscured AGNs, respectively. 
In Figure \ref{mex} we show the MEx diagnostic diagram for galaxies in the zCOSMOS survey with ASPIC spectral line measurements (grey points) with EW$ > $3 \AA~on the \oiiil, \oii~ and  \hb~ lines, which correspond to 5033 objects.

\begin{figure}
 	\includegraphics[width=100mm]{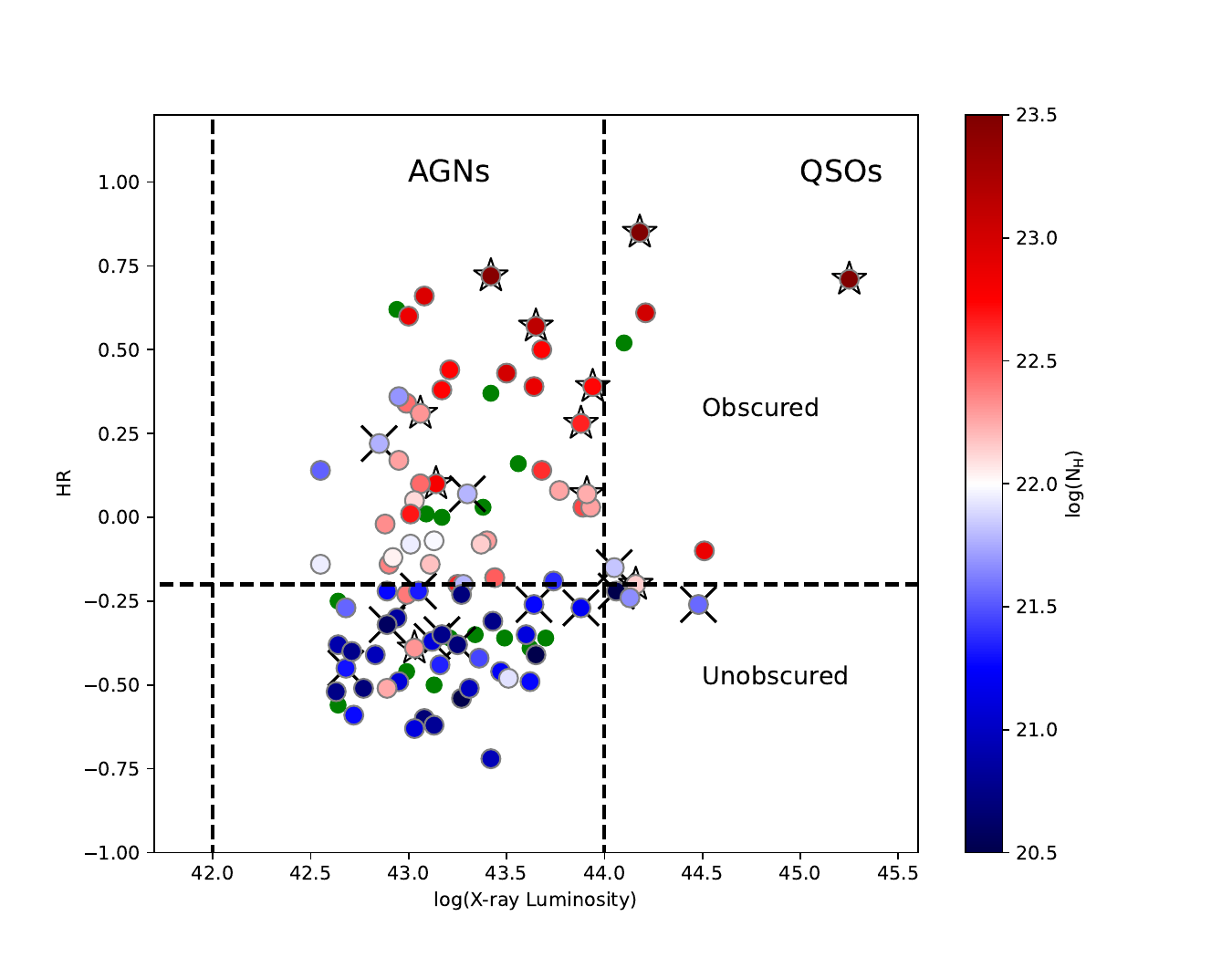}
   \caption{Hardness ratio as a function of hard (2-10 keV) X-ray luminosity. Vertical dashed lines show the typical separation for AGNs and quasars used in the X-rays. The horizontal line shows the limit values for obscured and unobscured AGNs in the X-rays. Stars and 'X' symbols indicate obscured and unobscured AGNs with high-excitation values (log(\oiiil/\hb)>0.5). Sources marked with green circles correspond to X-ray emitters found in the region of the MEx diagram that is generally associated with star-forming galaxies. Vertical colour bar shows the corresponding log($N_{\rm H}$) values.}
   \label{HR_Lx}
   \end{figure}

We have compared stellar masses from different methods in the COSMOS2015 and COSMOS2020 catalogues and found no significant differences. Thus, we adopted the COSMOS2015 MASS\_BEST values, derived from spectral fitting techniques.
Open circles represent sources with X-ray emission, while red and blue circles represent obscured and unobscured AGNs according to hydrogen column density estimates. Dashed lines show the separation criterion between AGNs, composites and star-forming galaxies taken from \citet{J14}.
We have also included the demarcation (horizontal dotted line at $y=0.5$) that allows the separation between high-and low-excitation AGNs from \citet{kewley2006}. 
We find 17 and 13 obscured and unobscured AGNs with high-excitation values (log(\oiiil/\hb)$\geq$0.5).

It is also noticeable that some X-ray emitting sources lie in the region of the MEx diagram typically populated by star-forming galaxies. A possible explanation for their X-ray emission could be the presence of high-mass X-ray binaries (HMXBs). \citet{mineo2012} analysed a sample of nearby galaxies hosting HMXB populations and found a correlation between the star formation rate (SFR) and the X-ray luminosity, described by the relation $L_X \approx 2.6 \times 10^{39} \times \mathrm{SFR} \quad \text{[erg s}^{-1}]$.
If the sources in the MEx diagram were powered solely by HMXBs, this would imply extremely high star formation rates, on the order of $10^3$--$10^4~M_\odot\,\mathrm{yr}^{-1}$, which are atypical for such systems (see Figure \ref{HR_Lx}). An alternative explanation is that these sources are optically dull AGNs-weakly accreting active nuclei with truncated accretion disks, likely associated with radiatively inefficient accretion flows \citep{Trump2009}.

\section{AGNs properties}
\label{properties}

\subsection{Redshift and stellar mass distributions}

We find that obscured and unobscured AGNs have similar redshift distributions, with mean values of $z_{\rm obs}=0.74\pm0.11$ and $z_{\rm unobs}=0.74\pm0.10$, respectively. Similarly, their stellar mass distributions are comparable, with mean values of $\log_{10}(\mathrm{mass}_{\rm obs})=10.7\pm0.3$ and $\log_{10}(\mathrm{mass}_{\rm unobs})=10.8\pm0.3$ (M$_\odot$). A Kolmogorov-Smirnov (K-S) null probability test confirms that the two samples of obscured and unobscured AGNs are statistically indistinguishable, with a null probability of $p_{\rm K\text{-}S}=0.18$. We did not differentiate between low- and high-excitation AGNs for this analysis.

\subsection{Hardness ratio and X-ray luminosity}

In this section, we examine the relationship between the X-ray hardness ratio and luminosity, providing insights into how absorption and intrinsic emission characteristics differentiate AGN populations.
Figure \ref{HR_Lx} presents the hardness ratio vs. the hard (2-10 keV) X-ray luminosity for obscured and unobscured AGNs. The vertical colour bar indicates the logarithm of the hydrogen column density (log($N_{\rm H}$).
Highly excited obscured and unobscured AGNs are marked with stars and X symbols, respectively. The vertical dashed line separates objects with log(X-ray)$>$44, typically associated with quasars, from those with log(X-ray)$<$44, which are more commonly classified as AGNs.

A clear trend emerges: high-excitation AGNs, both obscured and unobscured, tend to exhibit higher HR values compared to low-excitation AGNs.
Highly excited AGNs are often associated with relatively high accretion rates or powerful radiation fields, which generate strong optical emission lines, such as \oiiil. 
This intense activity also leads to significant X-ray emission in both the soft and hard bands, contributing to the high HR values observed.
For obscured AGNs, the surrounding gas and dust torus can absorb soft X-rays more effectively than hard X-rays, resulting in a higher HR value. However, the fact that even highly excited unobscured AGNs have large HR values suggests that their intrinsic X-ray emission is inherently hard. This could be linked to higher accretion rates or specific properties of the accretion flow around the black hole.

\subsection{Ionisation parameter versus X-ray luminosity}
\label{ioni}
\begin{figure}
	\includegraphics[width=85mm]{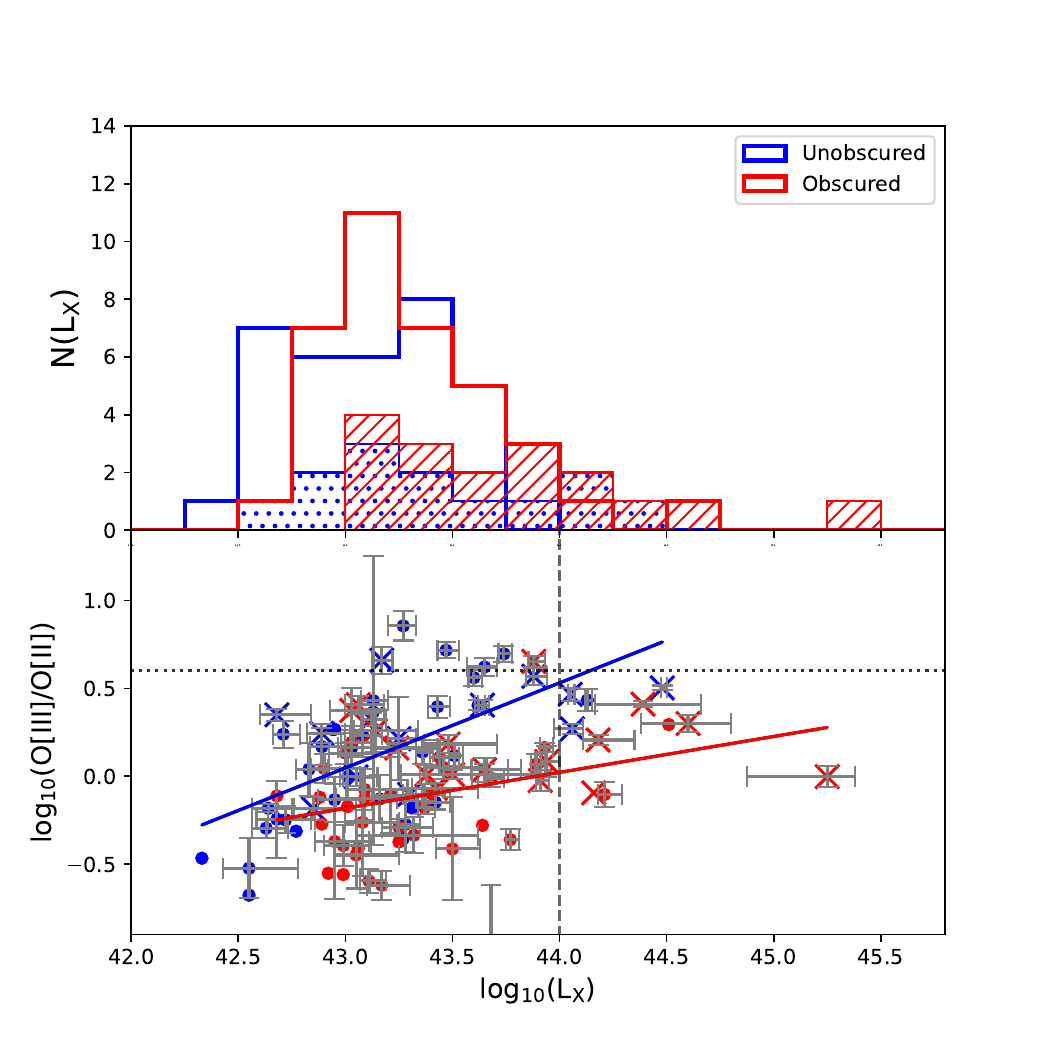}
   \caption{Ionisation-level sensitive \oiiil/\oii~ratio as a function of X-ray luminosity for the different AGN sample. Symbols are the same as Figure \ref{HR_Lx}. High-excitation obscured and unobscured AGNs are marked with 'X' symbols. Vertical dashed lines show the typical separation for AGNs and quasars used in the X-rays and the horizontal lines show the limit for sources with high-ionisation values.  }
    \label{o32lx}
\end{figure}

The ionisation state of AGNs is strongly influenced by their central energy output, making the relationship between the ionization-sensitive O32 ratio and X-ray luminosity a key probe of AGN activity. Investigating this connection allows us to explore how radiation from the accretion disk affects the surrounding narrow line region.
Figure \ref{o32lx} shows the ionisation-level sensitive O32 as a function of X-ray luminosity. 
AGNs with high-excitation values, both obscured and unobscured, are marked with 'X' symbols.
Vertical dashed lines indicate the typical separation between AGNs and quasars used in X-ray luminosity \citep{treister09}, while horizontal lines highlight sources with high-ionisation (O32$>$4) \citep{paalvast18}. We will address these high ionisation sources in great detail in the following sections.
The red and blue lines represent linear fits to the obscured and unobscured AGN samples, respectively, without distinguishing between low- and high-excitation AGNs.
We performed a linear regression analysis (without making any distinction between the different AGN types) to determine the relationship between the O32 parameter and X-ray luminosity: 
log(O32)=a$\times$[log(L$_{\rm X}$)$-$42.5]$+$b.

For obscured AGNs, we found a=(0.20$\pm$0.08), b=$-$0.28$\pm$0.08, and a Pearson coefficient R=0.341. For unobscured AGNs, the corresponding values are a=(0.48$\pm$0.09), b=($-$0.19$\pm$0.07) and a Pearson coefficient of 0.62.
These results indicate a stronger and more pronounced correlation between X-ray luminosity and ionisation parameter for unobscured AGNs, consistent with a less absorption-affected close central engine environment. In contrast, obscured AGNs exhibit a weaker relationship, suggesting that the absorbing material plays a significant role in modulating high-energy radiation and, consequently, the ionisation state of the surrounding gas.
The upper panel of Figure \ref{o32lx} shows the X-ray luminosity distributions for the different AGN samples. We find the following mean log(X-ray luminosity) values: obs\_low=43.30$\pm$0.40, obs\_high=43.80$\pm$0.60 and unobs\_low=43.10$\pm$0.40, unobs\_high=43.40$\pm$0.50. 
These results suggest that obscured, high-excitation AGNs tend to have higher X-ray luminosities on average. 

An alternative explanation for these spectral characteristics is that these objects are optically 'dull' AGNs, which are thought to host radiatively inefficient accretion flows (RIAFs; \citet{Trump2009}. These flows produce fewer ionising photons compared to standard accretion discs, leading to a reduced ionisation level in the narrow-line region. Consequently, this results in lower \oiiil/\oii~ratios \citep{Ho2008}, consistent with such AGNs harbouring inefficiently radiating central engines.

\subsection{Obscured and unobscured AGN according to the ionisation-level parameter}

\begin{figure*}
 	\includegraphics[width=85mm]{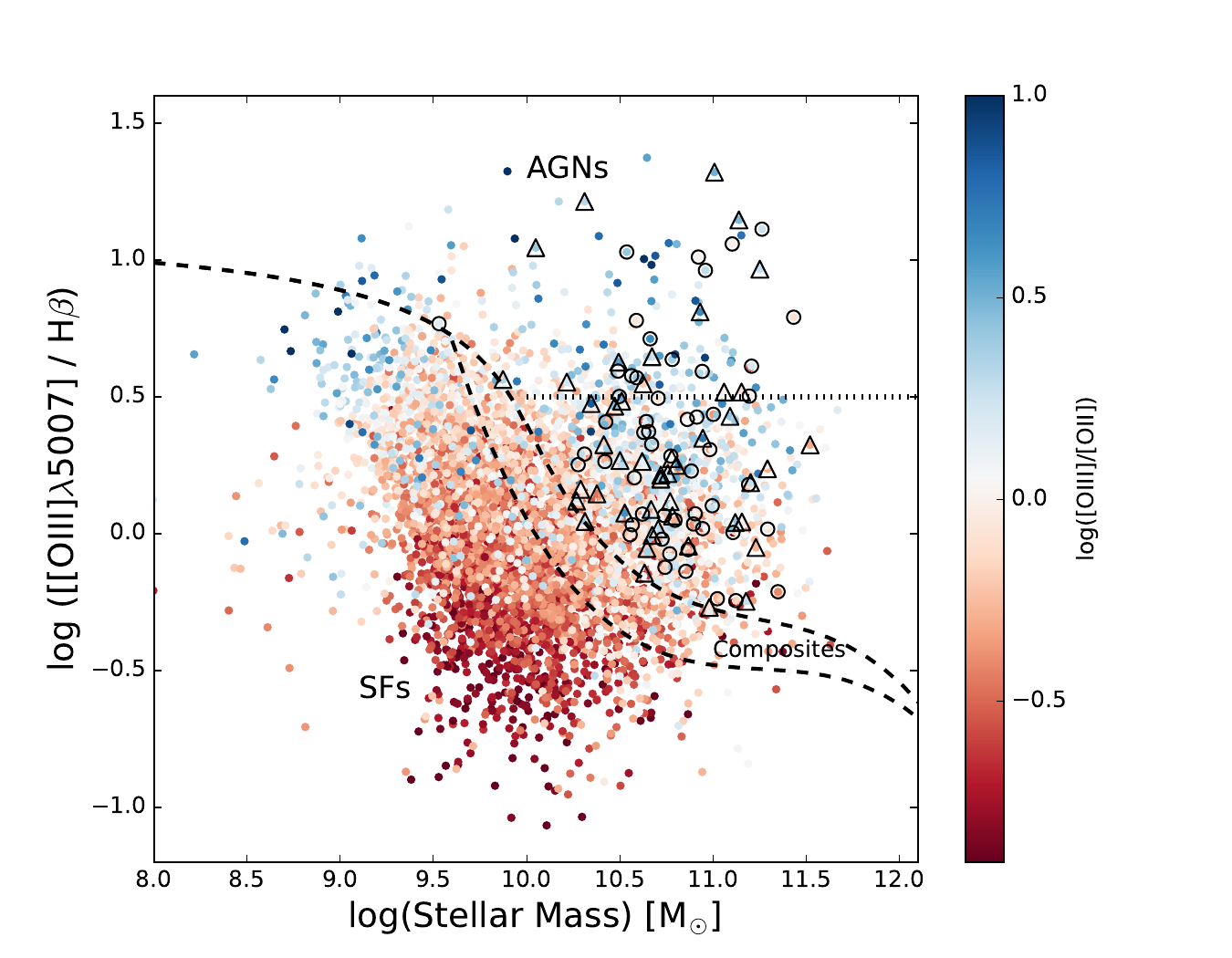}
	\includegraphics[width=85mm]{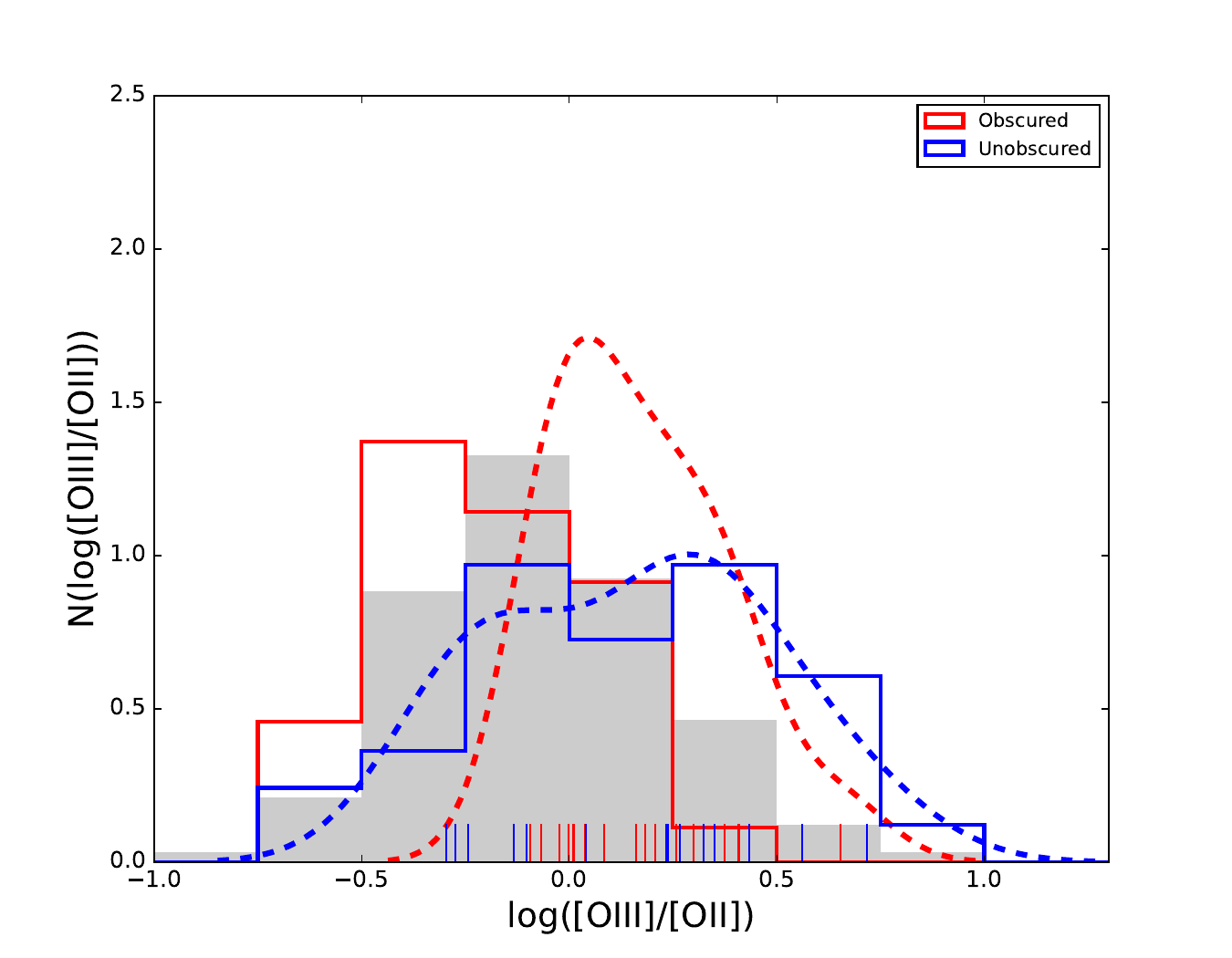}
   \caption{Left panel: Mass-Excitation diagram for the sample of galaxies in the zCOSMOS survey with ASPIC spectral line measurements. Obscured AGN with X-ray emission are plotted with open circles and the unobscured sample with open triangles. 
   The dashed and dotted lines demarcate the same regions as those in the Figure \ref{mex}.
   Vertical colour bar shows the ionisation-level \oiiil/\oii~ratio values. Right panel: ionisation-level \oiiil/\oii~ratio distributions for the different AGN samples. Red and blue solid line histogram represent the distribution for obscured and unobscured X-ray AGNs with low-excitation values.
   High-excitation AGNs are plotted with dashed line distributions that represent the kernel density estimates (KDEs) for obscured and unobscured X-ray AGNs. We plot also the individual observations with marginal ticks.}
    \label{MO32}
\end{figure*}

In this section we investigate the different AGN sample according to the ionisation-level O32 ratio.
Distinguishing between obscured and unobscured AGNs based on ionisation diagnostics provides deeper insight into the physical conditions of the narrow-line region (NLR). The O32 ratio serves as an important tracer of ionization levels, enabling a comparative analysis of AGNs with different levels of obscuration.
Figure \ref{MO32} (left panel) presents the MEx diagram for obscured (open circles) and unobscured (open triangles) AGNs. The vertical colour bar indicates the ionisation-level values.
In the right panel we have included the ionisation-level O32 ratio distributions for the different AGN samples.
Red and blue solid line histograms represent low-excitation obscured and unobscured X-ray AGNs (log(\oiiil/\hb$<$0.5), respectively. Red and blue dashed line histograms show the corresponding distributions for high-excitation AGNs (log(\oiiil/\hb~$\geq$0.5). The grey histogram represents MEx-selected AGNs without X-ray emission, matched in redshift and stellar mass to the X-ray AGN sample.
We find the following mean O32 values:
obs\_low$=$$-$0.21$\pm$0.27, obs\_high=0.17$\pm$0.20, unobs\_low=0.13$\pm$0.37, unobs\_high=0.15$\pm$0.31. 
For MEx selected AGNs without X-ray emission with similar redshift and stellar-mass values we find agn\_noXrays= $-$0.05$\pm$0.30.

\begin{figure*}
 	\includegraphics[width=85mm]{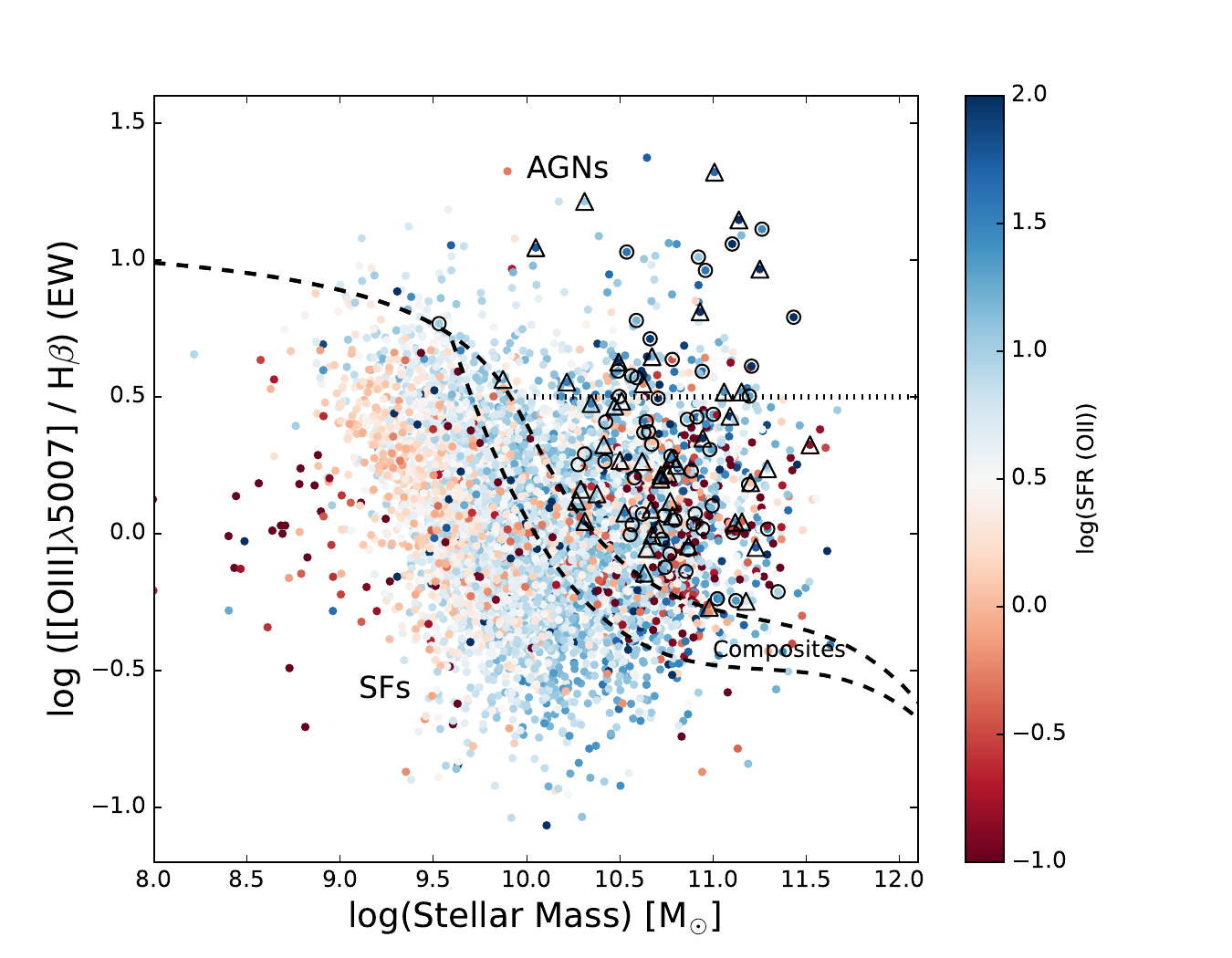}
   	\includegraphics[width=86mm]{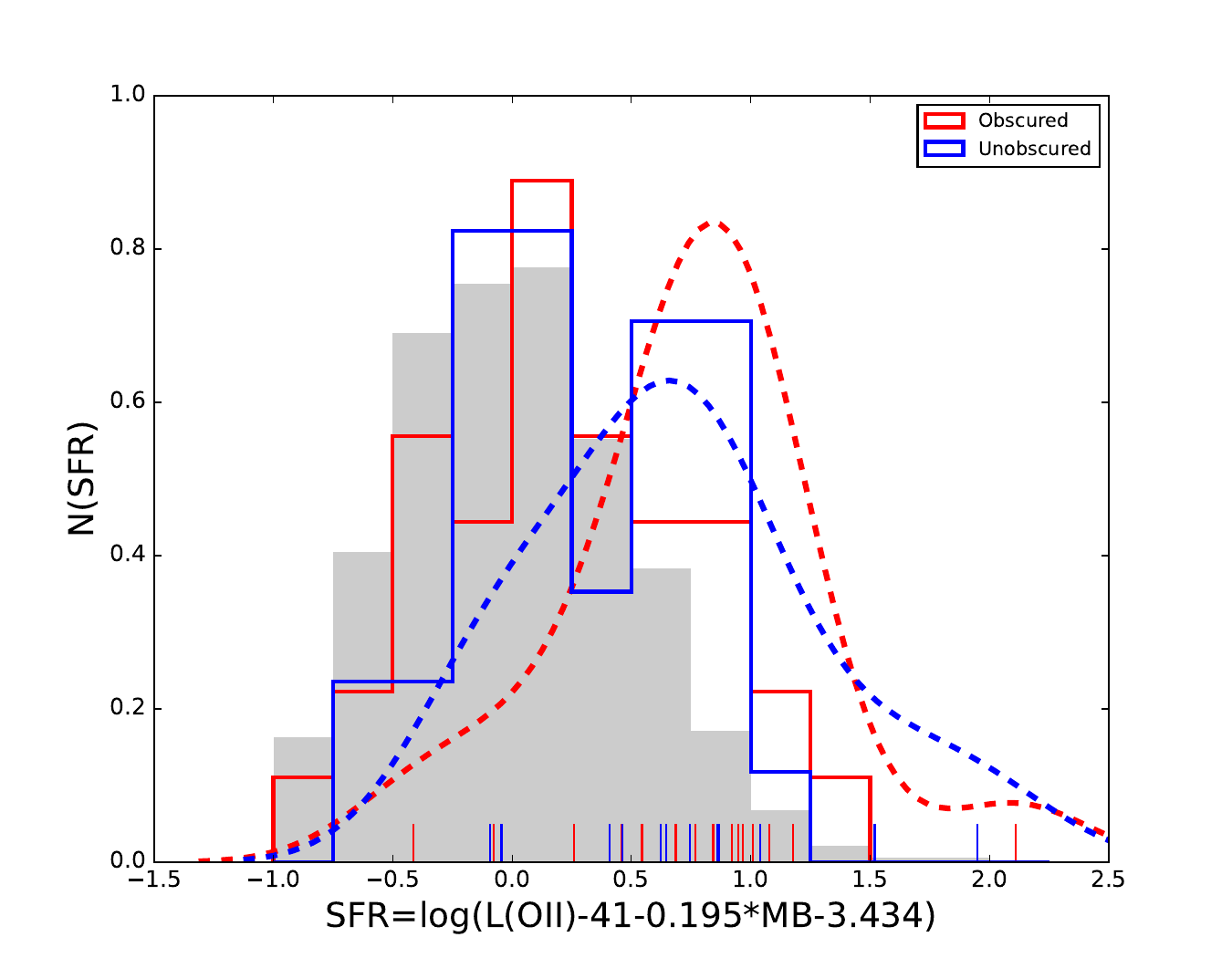}
   \caption{Left panel: MEx diagram for sources in the zCOSMOS and ASPIC catalogues. Vertical colour bar represent the SFR values obtained using the formula of \citet{maier09} from \oiil~luminosity and M$_B$ magnitudes. AGNs are represented as in previous figures. Right panel: SFR distribution for the different AGN samples. Colour code and histogram lines are the same as in previous figures. }
    \label{sfr}
\end{figure*}

High-excitation obscured and unobscured AGNs, along with low-excitation unobscured AGNs, exhibit similar O32 values (mean values $\sim$0.15 dex). However, low-excitation obscured AGNs show a significantly lower O32 value, differing by approximately 0.36 dex.
An analysis of variance (ANOVA; \citealt{fisher1925}) confirms that the mean differences across the four defined groups (obscured/unobscured × high/low excitation) are statistically significant ($p<10^{-7}$). Post-hoc comparisons using the Tukey honestly significant difference test \citet{tukey1949} show that the obscured low-excitation AGNs have systematically lower O32 values than the other three groups, which do not differ significantly among themselves. These results suggest that low-excitation obscured AGNs may represent a physically distinct population, possibly characterised by lower ionisation parameters or different narrow-line region conditions.
This is consistent with studies suggesting that obscured AGNs often experience a decrease in the ionisation parameter due to the presence of gas or dust in the near-nuclear environment \citep{dempsey18, Lu19}.
Alternatively, the observed trends may indicate that the ionization conditions in the NLR are not solely determined by AGN obscuration, but also reflect differences in host galaxy properties or AGN structure. The lack of a clear separation between obscured and unobscured AGNs in the high-excitation regime suggests that additional physical parameters such as gas density, metallicity, or ionizing photon flux may play a more dominant role in setting the ionization state in these systems.

\begin{figure*}
 	\includegraphics[width=85mm]{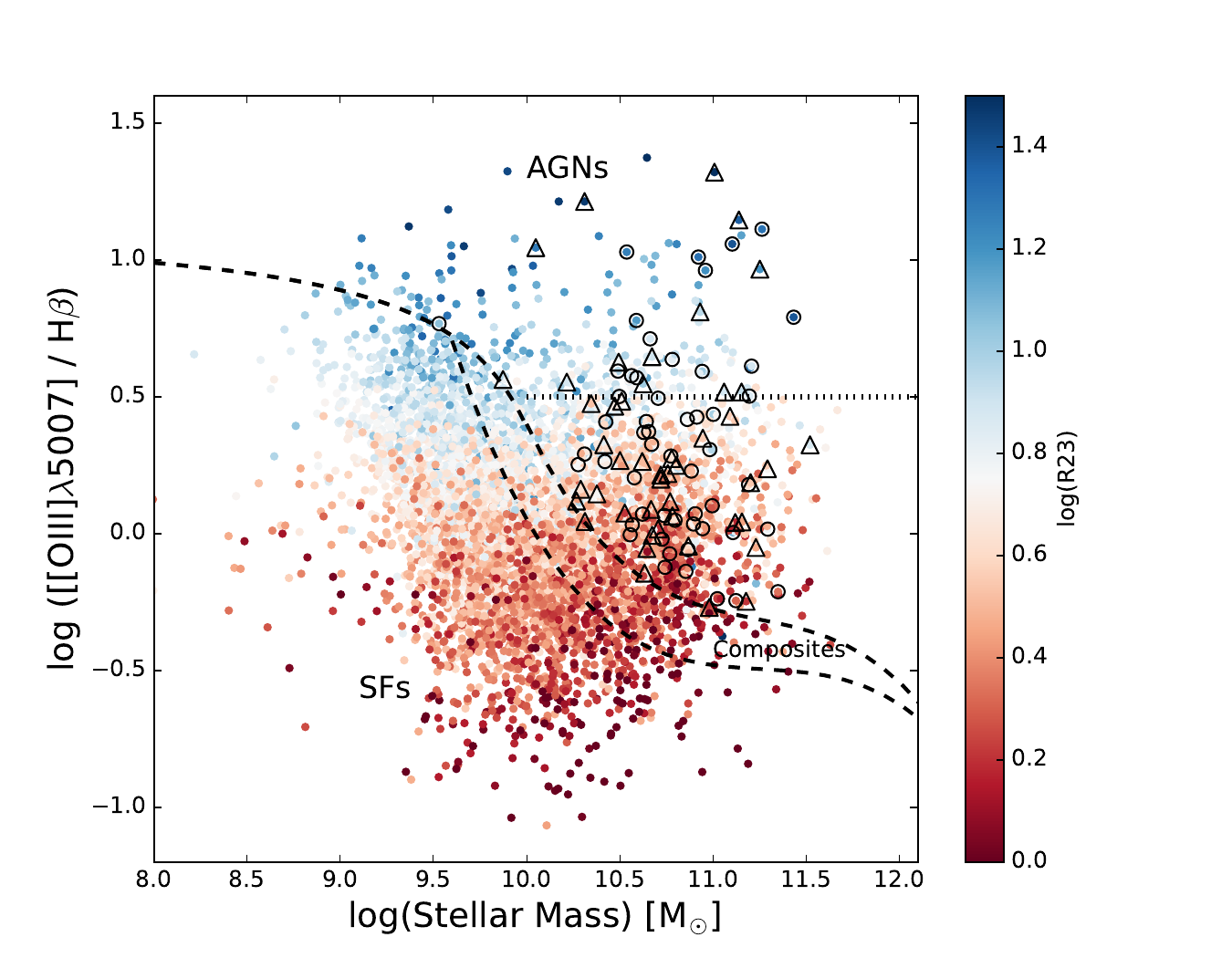}
	\includegraphics[width=85mm]{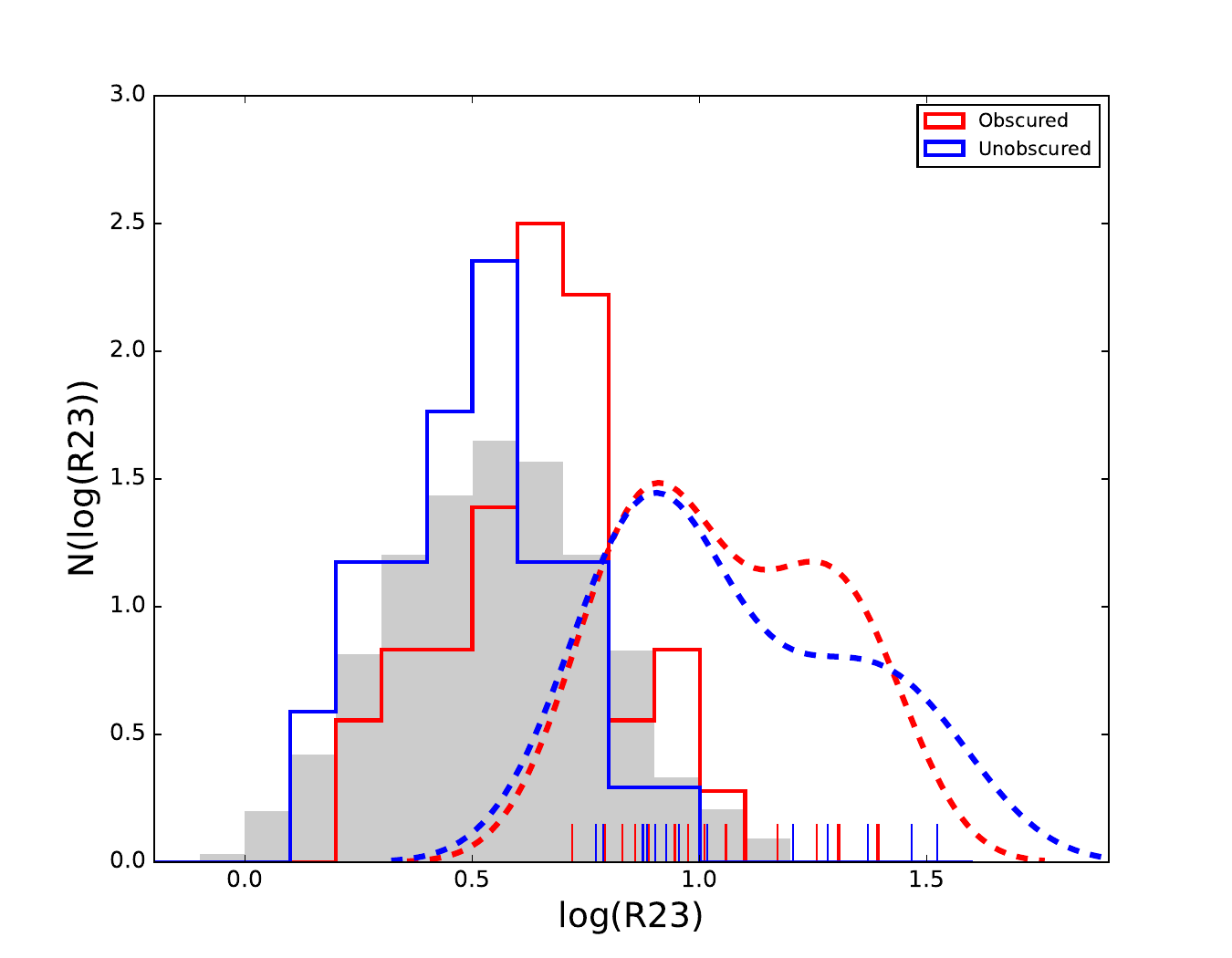}
   \caption{Left panel: MEx diagram for sources identified in zCOSMOS and ASPIC catalogues. Vertical colour bar shows the log(R23) parameter values. Right panel: Distribution of log(R23) values for the same AGN samples as in Figure \ref{MO32}.}
    \label{r23}
\end{figure*}

\begin{figure}
	\includegraphics[width=90mm]{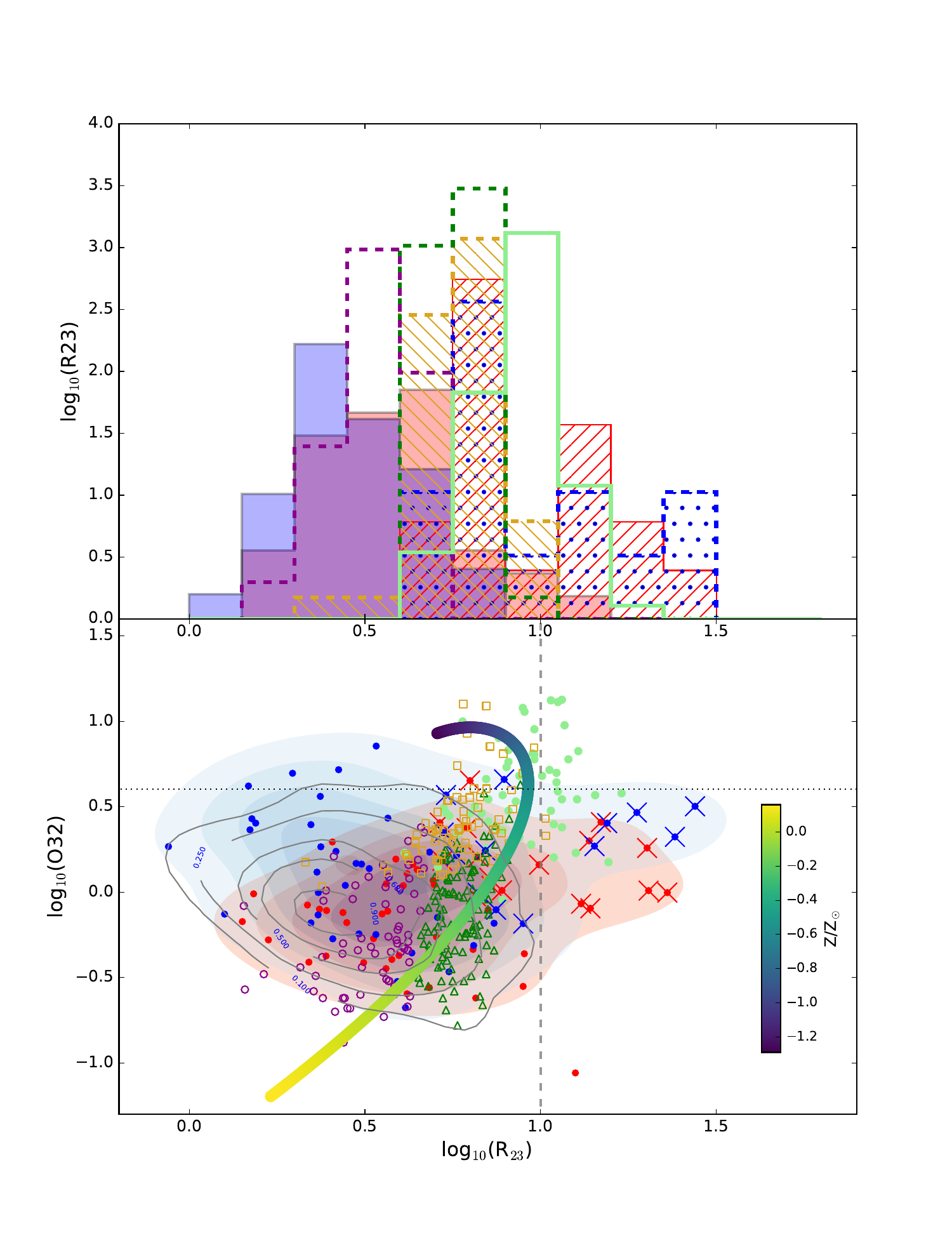}
	   \caption{Line ratio diagnostic diagram of R23=(\oiiil+\oii)/\hb~vs. O32=\oiiil/\oii). Obscured and unobscured AGNs are plotted with the same symbols as the previous figures. Purple open circles and dark green triangles represent galaxies with high (12+log(O/H)>8.8) and low (12+log(O/H)<8.8) metallicity taken from \citet{Kobul2004}. Golden open squares represent the corresponding values for extreme emission-line galaxies out to z$\sim$1 in zCOSMOS survey \citep{amorin15}. 
     Filled circles in light green colours represent high redshift galaxies ($4<z<6$) with 12+log(O/H) $<$ 8.5 from JWST/NIRSpec survey \citep{nakajima2023}.
      Grey contours show the corresponding distribution for MEx selected AGN without X-ray emission with the same redshift and stellar-mass limits as our AGN sample.
    The strong line diagnostic curve indicates the colour coded metallicity values in solar units (Z/Z$_{\odot}$) taken from \citet{curti17}.}
       \label{o23r23}
\end{figure}

\subsection{Star Formation Rate}

The interplay between AGN activity and star formation is a fundamental aspect of galaxy evolution. By examining the star formation rates (SFRs) of AGN host galaxies, we can investigate whether AGN-driven feedback influences ongoing star formation, particularly in the context of obscured and unobscured sources. We calculated the star formation rate (SFR) using the estimator proposed by \citet{maier09}:

\begin{equation}
\label{oiisfr}
\rm{log}[SFR/M_{\odot}\,yr^{-1}]=\rm{log}[L_{[OII]}/(ergs\,s^{-1})]- 41  - 0.195 \times M_{B} - 3.434
\end{equation}

where L\oiil~is the luminosity of the \oii~line and M$_{\rm B}$ is the absolute $B$ band magnitude.
Figure \ref{sfr} (left panel) presents the MEx diagram for the sample of galaxies and AGNs. The vertical colour bar indicates the SFR values calculated using equation \ref{oiisfr}. The right panel shows the SFR distribution for the AGN sample.   
We find the following mean values: obs\_low=0.24$\pm$0.53, unobs\_low=0.28$\pm$0.46, obs\_high=0.75$\pm$0.52, unobs\_high=0.68$\pm$0.57 and the sample of AGN selected according with the MEx diagram without X-ray emission, noX=$-$0.015$\pm$0.5.
As shown, both obscured and unobscured AGNs with high-excitation values exhibit higher SFRs compared to low-excitation AGNs. Objects with high O32 values ($>$ 0.6), as shown in Figure \ref{o32lx}, also tend to have high SFRs (with values of the order of 1 calculated using the formula \ref{oiisfr}), suggesting a connection between high ionisation and active star formation in these AGNs. 
We performed Welch’s t-test \citep{welch1947} to compare the distributions between obscured and unobscured sources within the low- and high-excitation groups. Welch’s test was chosen because it accounts for unequal variances and sample sizes between groups, providing a more reliable comparison under these conditions. No statistically significant differences were found between obscured and unobscured samples in either excitation regime ($p > 0.7$). In contrast, significant differences emerge when comparing low- versus high-excitation sources within both obscured ($p = 0.0023$) and unobscured ($p = 0.036$) groups. These findings indicate that the excitation level, rather than obscuration, is the main factor influencing the observed variations in the analysed parameter.

Similar results were published by \citet{Nakajima} in a sample of galaxies with z=2-3 who found that galaxies with large O32 values are associated with less massive galaxies with efficient star formation.

\subsection{O32 and R23 parameters}

\citet{pagel79} first introduced the R23 parameter, defined as R23=(\oiiil+\oii)/\hb~to study the properties of HII regions in nearby galaxies. This parameter has since been widely used in the literature. R23 is sensitive to metallicity, exhibiting a biphasic behavior.
For low-metallicity galaxies, R23 increases with metallicity until it reaches a maximum at slightly less than solar abundance \citep{curti17, Nakajima2022}. However, for high-metallicity galaxies, R23 decreases as the gas phase O/H ratio increases. This occurs because oxygen, an efficient coolant, reduces the gas temperature, leading to fewer collisionally excited oxygen ions and weaker oxygen emission lines \citep{perrotta21}. This phenomenon, known as the "double-value degeneracy" \citep{curti2020}, means that the same R23 value can be obtained for two different metallicity values.

\begin{figure*}
	\includegraphics[width=150mm]{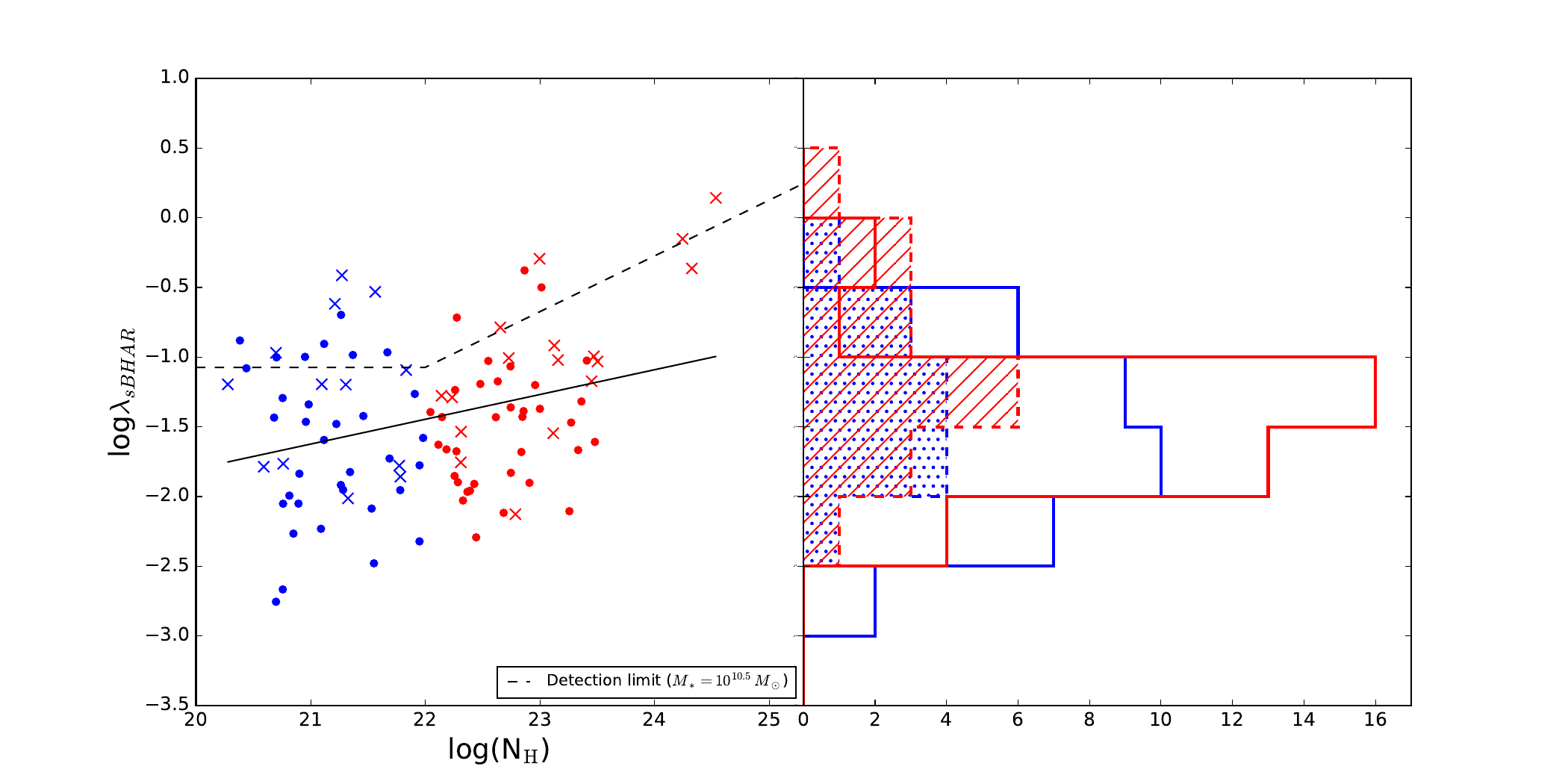}
	   \caption{Specific black hole accretion rate \Sar\ as a function of hydrogen column density, log($N_{\rm H}$) for obscured and unobscured AGNs. Low (high) excitation obscured and unobscured AGNs are plotted with solid (X's) red and blue circles. The dashed black line indicates the minimum detectable $\lambda_{\mathrm{sBHAR}}$, as a function of $\log(N_{\mathrm{H}})$, for a fixed stellar mass of $\log(M_{*}/M_{\odot}) = 10.5$ and a mean redshift of $z = 0.75$. This detection limit accounts for the hard-band (2--10 keV) flux threshold and includes attenuation due to obscuration, estimated using an empirical transmission model. Sources lying below this line are likely to fall below the detection limit of the X-ray survey.    
       In the left panel we plot the corresponding distribution of \Sar\ for the different AGN samples. Solid red and blue distributions represent obscured and unobscured AGNs, while red dashed and blue dotted line histograms correspond to high-excitation obscured and unobscured AGNs, respectively.
} 
    \label{lambdaNH}
\end{figure*}

 \begin{figure*}
 	\includegraphics[width=85mm]{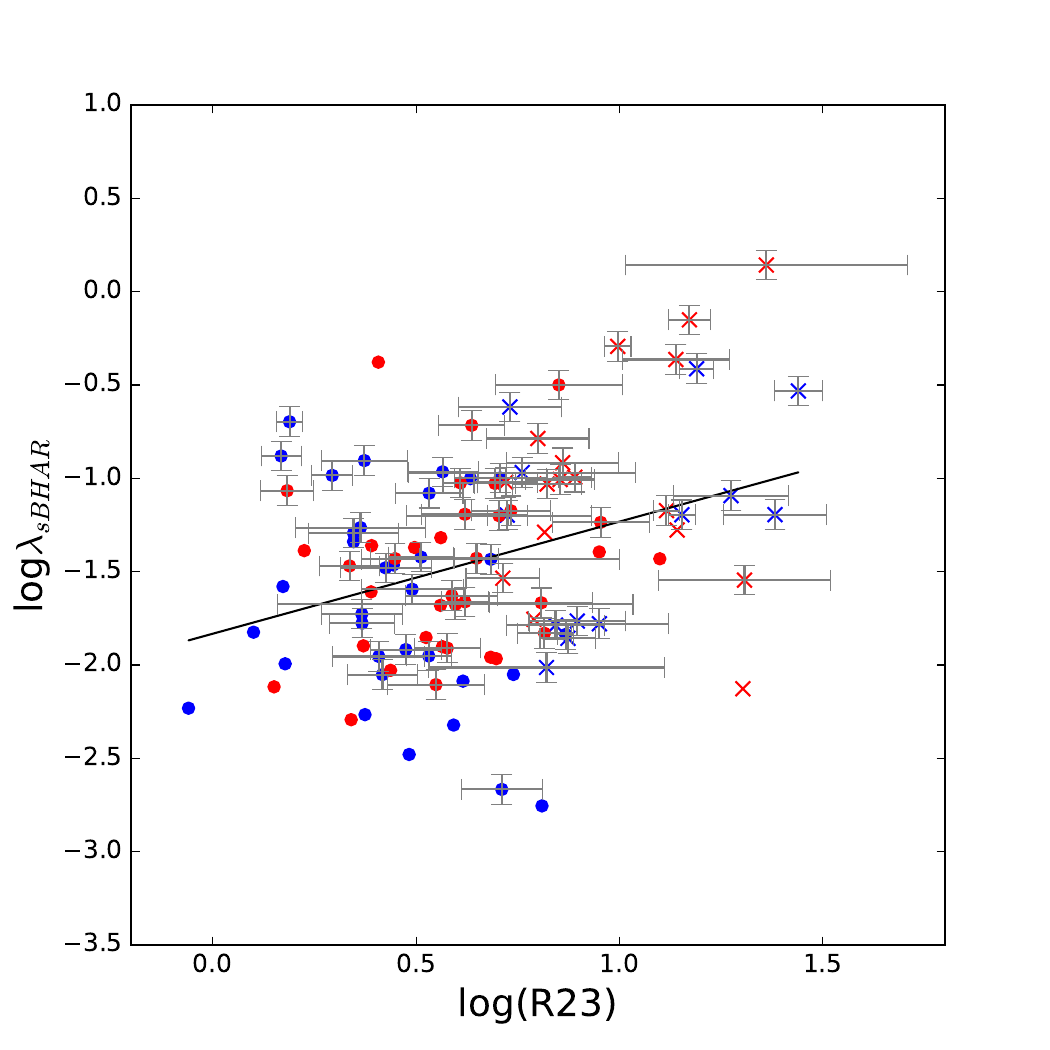}
   	\includegraphics[width=85mm]{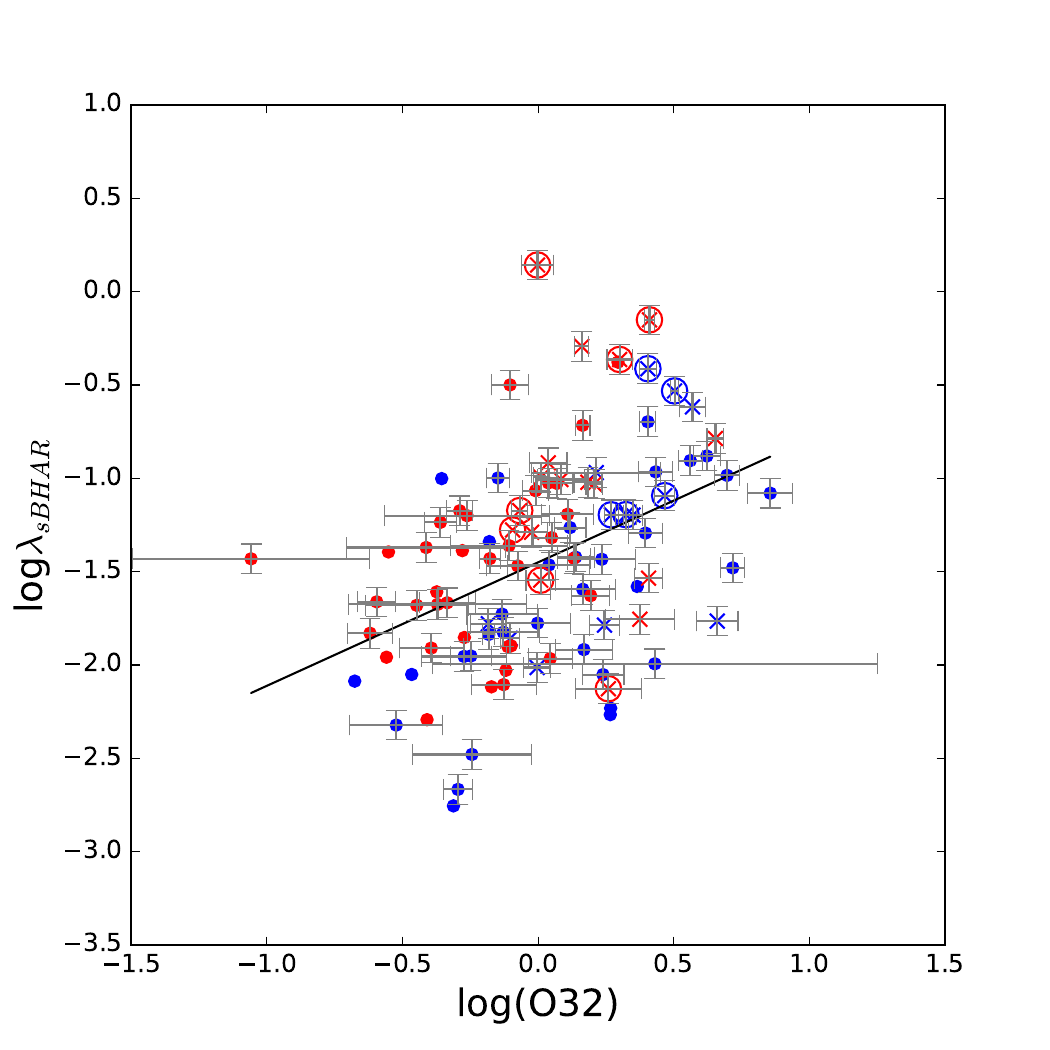}
	   \caption{Specific black hole accretion rate \Sar\ as a function of R23 (left panel) and O32 (right panel) parameter for the different AGN samples.
    The plotted symbols are the same as those in Figure \ref{lambdaNH}. Objects without error bars correspond to sources for which the uncertainty in the H$\beta$ emission line is not available. Points enclosed by large circles represent sources with log(R23)>1. }
    \label{lambdaR23}
\end{figure*}

Figure \ref{r23} shows the MEx diagram (left panel) and the corresponding R23 distribution (right panel) for the AGN samples.
Low-excitation obscured AGNs exhibit higher R23 values (0.64-0.5=0.14 dex) compared to the unobscured AGNs. Both obscured and unobscured high-excitation AGNs show similar R23 distributions (around 1.05 and 1.07 dex).
A high R23 value often indicates a strong ionising radiation field in the AGNs surrounding region. In such environments, oxygen is in higher ionisation states (\oiii~and \oiil), suggesting that the AGN's radiation is energetic enough to efficiently ionise the gas. This is typical of AGNs with a highly active accretion disk or powerful central source.
In some cases, high R23 values are associated with low metallicity in the surrounding gas. In low-metallicity regions, fewer heavy metals are available to cool the gas radiatively, allowing the ionised oxygen lines to be more prominent relative to \hb. 
We performed Welch’s t-tests \citep{welch1947} to compare the analysed parameter between different groups, accounting for unequal variances and sample sizes. Significant differences were found when comparing high- versus low-excitation sources within both obscured ($t = 6.76$, $p < 0.001$) and unobscured ($t = 7.37$, $p < 0.001$) samples. Additionally, a significant difference was observed between obscured and unobscured sources in the low-excitation regime ($t = 2.85$, $p = 0.006$). In contrast, no statistically significant difference was found between obscured and unobscured sources in the high-excitation group ($t = -0.19$, $p = 0.85$). These results reinforce the conclusion that excitation level predominantly influences the observed variations, with obscuration playing a less significant role, particularly in the high-excitation regime.

AGNs with high R23 values may reside in environments with lower metallicity, possibly due to gas accretion from the circumgalactic or intergalactic medium.
In general, galaxies with high R23 and log(\oiiil/\hb)$>$0.5 are indicative of metal-poor environments and significant stellar activity, making them intriguing objects for studying galactic evolution and star formation processes.

The O32 vs. R23 diagram is a useful tool for investigating metallicity and ionisation state in the local universe and up to $z=2-3$ \citep{kewley02, Nakajima}.
In Figure \ref{o23r23} shows the O32 parameter plotted against the R23 parameter for obscured and unobscured AGNs. High-excitation AGNs are marked with 'X' symbols. We also plot the corresponding values of extreme emission-line galaxies (z$<$0.9) from the zCOSMOS survey \citep{amorin15} (golden open squares). 
We have correlated the Amorin et al. sample of galaxies with the spectral line measurements from the ASPIC catalogue and have taken into account the same definitions used in our AGN samples (0.5$<z<$0.9 and same cuts in the equivalent widths in the \oiiil, \oii and \hb~ lines).
Open circles and triangles in purple and dark green shows the distribution of galaxies with 0.3$<z<$1 in the
Great Observatories Origins Deep Survey–North (GOODS-N) field with high (12+log(O/H)>8.8) and low (8.5<12+log(O/H)<8.8) metallicity values \citep{Kobul2004}. 
We plot also, a sample of high-redshift galaxies ($4<z<6$) with 12+log(O/H) $<$ 8.5, identify in the JWST/NIRSpec survey (filled circles in lightgreen colour,   \citet{nakajima2023})
Grey contours show the corresponding distribution for MEx selected AGN without X-ray emission with the same redshift and stellar-mass limits as our AGN sample.
A statistical analysis using Student’s t-test reveals that the values of log(R23) show a significant difference between the samples of AGNs without X-ray emission and the obscured and unobscured AGNs. In contrast, according to the same statistical analysis, the values of log(O32) between the different samples do not show a significant difference, and the null hypothesis of equal means is not rejected.
A group of high-excitation obscured and unobscured AGNs exhibit high R23 values (log(R23)>1) and relative low O32 values. This is indicative of galaxies in more advanced stages of chemical evolution, where there is sufficient oxygen present, but the ionising radiation is not energetic enough to sustain a high proportion of doubly ionised oxygen.
In summary, this combination suggests that galaxies, although oxygen-rich and high R23, possibly originating in actively forming stars \citep{curti17, jiang2019}, do not experience extreme ionisation conditions, resulting in moderate to low O32 values, likely due to lower metallicity or less intense UV radiation.

\subsection{Specific black hole accretion rates $\lambda_\mathrm{sBHAR}$, obscuration and metallicity }
\label{lambda}

In the context of galactic evolution, SMBHs play a crucial role, particularly during active growth phases characterised by the accretion of material from their surroundings. This accretion process, along with the emission of large amounts of electromagnetic radiation from the accretion disk, is a characteristic feature of AGNs \citep{Yang2018}.

A key parameter for understanding the relationship between SMBHs and their host galaxies is the specific black hole accretion rate ($\lambda_\mathrm{sBHAR}$). This parameter, which quantifies the ratio of the black hole accretion rate to the stellar mass of the galaxy, allows us to explore how accretion processes relate to the global properties of host galaxies \citep{lopez23}. More specifically, $\lambda_\mathrm{sBHAR}$ provides insight into the efficiency with which SMBHs consume material relative to the total stellar mass of the galaxy \citep{mountrichas22, mountrichas24}.

We define \Sar\ in dimensionless units as:
\begin{equation}
\sar  = \frac{ k_\mathrm{bol} \; L_\mathrm{X} }
                  { 1.3 \times 10^{38} \; \mathrm{erg\;s^{-1}} \times 0.002 \dfrac{\mathcal{M}_*}{\mathcal{M}_\odot} }
\label{eq:sar}
\end{equation}
where $L_\mathrm{X}$ is the rest-frame 2-10 keV X-ray luminosity,
$k_\mathrm{bol}$ is a bolometric correction factor (we adopt a constant $k_\mathrm{bol}=25$, as in \citet{Aird18, mountrichas22, mountrichas24}, and $\mathcal{M}_*$ is the total stellar mass of the AGN host galaxy estimated from SED fitting. 
The significance of $\lambda_\mathrm{sBHAR}$ lies in its ability to directly compare accretion activity between galaxies of different masses, mitigating biases related to galaxy size or total stellar mass. A high $\lambda_\mathrm{sBHAR}$ indicates that the black hole is in an active growth phase, often associated with the luminous AGNs, while a low value suggests lower accretion, typical of obscured AGNs or less active phases \citep{mountrichas24}.

Figure \ref{lambdaNH} illustrates the correlation between $\lambda_\mathrm{sBHAR}$ and the hydrogen column density, $N_{\rm H}$.
Filled red and blue circles represent low-excitation obscured and unobscured AGNs, respectively. AGNs with high-excitation values are marked with 'X' symbols (left panel). In the right panel we plot the distribution of \Sar\ for the different AGN samples.
We find the following mean log($\lambda_\mathrm{sBHAR}$) values: 
un\_low$=-$1.65, un\_high$=-$1.26, obs\_low$=-$1.49 and obs\_high$=-$1.01.
We find a weak but positive trend line between log($\lambda_\mathrm{sBHAR}$) and log($N_{\rm H}$). 
The following linear regression was obtained without differentiating between the various AGN types:
log($\lambda_\mathrm{sBHAR}$)=(0.18$\pm$0.05)$\times$(log$(N_{\rm H})-20)-$(1.8$\pm$0.1). However, the weak Pearson correlation coefficient (R=0.31) suggests a limited direct dependence between the two variables.
It would seem that the trend might be driven primarily by three Compton-thick AGN (log$(N_{\rm H} >$ 24). For this reason we have evaluated the influence of the three Compton-thick AGNs. When we exclude the three Compton-thick sources, the correlation weakens significantly to $R = 0.16$, and the regression becomes:
\[
\log(\lambda_{\mathrm{sBHAR}}) = (0.093 \pm 0.06)\, \log(N_{\mathrm{H}}) + (-1.67 \pm 0.13).
\]
These results indicate that a small number of highly obscured sources may drive the apparent positive trend, which becomes less significant when they are removed. Nevertheless, the slope remains positive, suggesting a possible, although tentative, connection between obscuration and black hole accretion activity.
To further assess the impact of selection effects, we estimated the minimum detectable $\lambda_{\mathrm{sBHAR}}$ as a function of $N_{\mathrm{H}}$, assuming a stellar mass of $\log(M_*/M_{\odot}) = 10.5$ and a redshift of $z = 0.75$, which are representative values for our sample. This limit, indicated by a dashed line in Figure~9, incorporates an attenuation correction derived from the torus model \citep{Murphy2009} and the observed hard-band flux limit. We find that most Compton-thick AGNs ($\log(N_{\mathrm{H}}) > 24$) lie near this threshold, suggesting that some heavily obscured sources with lower accretion rates may fall below the detection limit. However, the detection of several such sources above this boundary indicates that the observed positive trend between $\lambda_{\mathrm{sBHAR}}$ and $N_{\mathrm{H}}$ cannot be entirely attributed to selection bias.

It is reasonable to hypothesise a negative correlation between obscuration (as measured by $N_{\rm H}$) and the black hole accretion rate, meaning that less obscured AGNs should exhibit higher accretion rates. This expectation is consistent with the findings of \citet{Vijarnwannaluk2024}, who compared $N_{\rm H}$ values with the Eddington ratio, which is often used as a proxy for \(\lambda_{\mathrm{BHAR}}\) \citep{lopez23,mountrichas24}. Similar results were obtained by \citet{laloux2024}, who found that unobscured AGNs exhibit a systematic offset towards higher Eddington ratio compared to their obscured counterparts.
However, a positive correlation between these parameters has also been reported in some studies, such as those by \citet{Lanzuisi2015, Barchiesi2024}.
A higher accretion rate may be associated with an increased hydrogen column density, particularly if the surrounding material is denser or located closer to the black hole. This could result in stronger obscuration, although the exact nature of this relationship may depend on the AGN geometry, the distribution of the surrounding material, and other factors \citep{buchner2015}.

Figure \ref{lambdaR23}, left panel, plots $\lambda_\mathrm{sBHAR}$ against the R23 parameter. Both obscured and unobscured high-excitation AGNs (marked with 'X' symbols) exhibit lower metallicities, as indicated by the R23 parameter. We find a weak (or marginal) positive correlation between these two quantities although the scatter remains significant. We obtained the following linear regression (considering all AGN types together) for the relationship between $\lambda_\mathrm{sBHAR}$ and the R23 parameter: log$(\lambda_\mathrm{sBHAR})$=$0.60\times$log(R23)$-$1.83 with a Pearson coefficient R=0.34.

The correlation between the R23 parameter and $\lambda_\mathrm{sBHAR}$ may provide insights into how black hole accretion activity is influenced by the chemical environment of the host galaxy. In general, higher metallicity is expected to be associated with a lower specific accretion rate, as more metal-rich galaxies tend to have less gas available for accretion. 
Although, for example, \citet{du2014} did not find a correlation between metallicity-estimated through the ratio of certain emission lines and the accretion rate of black holes, these authors found that neither the BLR nor the NLR metallicity correlates with black hole masses or Eddington ratios. However, they reported a strong correlation between NLR and BLR metallicities.

In the right panel of Figure \ref{lambdaR23} we plot the $\lambda_\mathrm{sBHAR}$ vs. O32 ionisation-level parameter. 
As it can be seen we find a weak but positive correlation between these two parameters. 
The linear regression is the following: log$(\lambda_\mathrm{sBHAR})$=$0.66\times$log(O32)$-$1.45 and a Pearson correlation coefficient R=0.42. As in previous cases, we have not made any distinction between the different AGN types. We have marked with open circles sources with log(R23)>1. These objects are above the linear relationship, indicating that on average they have large black hole accretion rates.
No significant trend was observed when comparing obscured and unobscured AGNs as well as high- and low-excitation sources in our sample. 
This may be because the O32 parameter measures the ionisation in the narrow-line region, which is located on a larger scale than the obscuring region of the torus. If the ionising radiation from the AGN can escape in a similar manner in both cases (obscured and unobscured), then the O32 value would not be affected by the obscuration measured from $N_{\rm H}$.

\section{Summary and discussions}
\label{sum}

In this study, we analysed the properties of X-ray selected AGNs 
based on the Mass-Excitation (MEx) diagram. We categorised AGNs into obscured and unobscured sources using hydrogen column density estimates (log($N_{\rm H}$)=22) and into high- and low-excitation regimens using the \oiiil/\hb~ ratio of 0.5 as a threshold. 
We investigated various AGN properties, including X-ray luminosity, stellar mass, hardness ratio, specific black hole accretion rate ($\lambda_\mathrm{sBHAR}$), and emission line ratios like \oiiil/\oii~and R23.

Our main findings are as follows:

\begin{itemize} 

\item High-excitation AGNs: both obscured and unobscured high-excitation AGNs exhibit larger hardness ratios compared to low-excitation AGNs. This suggests that these AGNs are associated with more active nuclei and higher accretion rates, where the surrounding gas plays a crucial role in both X-ray absorption and optical line excitation.

\item Ionisation and X-ray luminosity: unobscured AGNs show a stronger correlation between the \oiiil/\oii~ratio and X-ray luminosity compared to obscured AGNs. Additionally, high-excitation obscured AGNs tend to have higher X-ray luminosities on average. These findings suggest that obscured and highly excited AGNs are associated with more active nuclei or higher accretion rates.

\item Ionisation levels: high-excitation obscured and unobscured AGNs, along with low-excitation unobscured AGNs, have similar \oiiil/\oii~ratios. However, low-excitation obscured AGNs exhibit significantly lower values, potentially due to obscuration blocking ionising radiation from the central engine or different physical conditions in the broad-line region.

\item Metallicity and ionisation: both highly excited obscured and unobscured AGNs tend to have larger R23 values, indicating lower metallicities in their host galaxies. A fraction of high-excitation AGNs exhibit extremely low metallicity values, similar to high-redshift galaxies observed with JWST.

\item Accretion rate and obscuration: we find a positive correlation between $\lambda_\mathrm{sBHAR}$ and $N_{\rm H}$, suggesting that obscuration may be related to specific phases of rapid black hole growth. However, there is no clear relationship between high-excitation obscured and unobscured sources, implying that AGNs in active accretion phases may undergo episodes of obscuration and unobscuration. 

\item Accretion, metallicity and ionisation: AGNs with high R23 values and high excitation are associated with low-metallicity environments. Higher metallicity is expected to correspond to a lower specific accretion rate, as metal-rich galaxies typically have less gas available for accretion.

\bigskip

These findings provide valuable insights into the ionisation properties and environmental conditions of AGNs, shedding light on the physics of active nuclei under various obscuration and excitation scenarios. The results highlight the importance of the surrounding environment in modulating the observed properties of AGNs and suggest that obscuration and excitation are closely linked to physical processes like X-ray absorption, accretion rate, and the chemical composition of the medium.

The observed differences between various types of AGNs, including obscured and unobscured sources, as well as those with high or low excitation, along with their properties such as metallicity, ionisation state, and black hole accretion rate, can be explained as deviations within the unified models. In this context, the size of the torus, in particular, may play a decisive role in determining these classifications.
Several studies have proposed modifications to the unified model of AGNs. For instance, \citet{gould12} analysed a sample of heavily obscured local AGNs- specifically, Compton-thick sources with \(N_H > 1.5 \times 10^{24} \,\mathrm{cm^{-2}}\) and suggested that the primary contributor to the observed mid-IR dust extinction is not the torus, but rather dust located at much larger scales within the host galaxy. This interpretation aligns with previous findings by other authors \citep{malkan98,Matt2000,Guainazzi2001, Goulding2009,sokol2023}.

Our findings support an evolutionary scenario where obscured sources accrete more rapidly due to the availability of gas in their surroundings \citep{dimatteo2005, sanders07, hopkins08a, hopkins08b, ricci21}. These studies are crucial for understanding the complex interplay between AGNs and their host galaxies, as well as the feedback processes that regulate star formation and galaxy evolution in the Universe \citep{hickox2018, remetallica}.

\end{itemize}

\begin{acknowledgements}
We thank the anonymous referee for his/her useful comments and suggestions.

This work was partially supported by the Consejo Nacional de Investigaciones Cient\'{\i}ficas y T\'ecnicas (CONICET) and the Secretar\'ia de Ciencia y Tecnolog\'ia de la Universidad Nacional de C\'ordoba (SeCyT). 
Based on data products from observations made with ESO Telescopes at the La
Silla Paranal Observatory under ESO programme ID 179.A-2005 and on
data products produced by TERAPIX and the Cambridge Astronomy Survey
Unit on behalf of the UltraVISTA consortium.
Based on zCOSMOS observations carried out using the Very Large Telescope at the ESO Paranal 
Observatory under Programme ID: LP175.A-0839 (zCOSMOS).
This research has made use of the VizieR catalogue access tool, CDS,  Strasbourg, France (DOI : 10.26093/cds/vizier). 
This research has made use of the ASPIC database, operated at CeSAM/LAM, Marseille, France.\\

\end{acknowledgements}

\begin{appendix}

\end{appendix}

\end{document}